\documentclass[12pt,preprint]{aastex}
\begin{document}

\shorttitle{STRUCTURE OF THE LOCAL INTERSTELLAR MEDIUM. V.}
\shortauthors{Redfield \& Falcon}

\title{THE STRUCTURE OF THE LOCAL INTERSTELLAR MEDIUM V: ELECTRON DENSITIES}

\author{Seth Redfield\altaffilmark{1} and Ross E. Falcon}
\altaffiltext{1}{Hubble Fellow}
\affil{\vspace{-2mm}Department of Astronomy and McDonald Observatory, University of Texas, Austin, TX, 78712}
\email{\vspace{-3mm}sredfield@astro.as.utexas.edu; cylver@astro.as.utexas.edu}

\begin{abstract}

We present a comprehensive survey of \ion{C}{2}$^{\ast}$ absorption
detections toward stars within 100~pc in order to measure the
distribution of electron densities present in the local interstellar
medium (LISM).  Using high spectral resolution observations of nearby
stars obtained by the Goddard High-Resolution Spectrograph (GHRS) and
the Space Telescope Imaging Spectrograph (STIS) onboard the {\it
Hubble Space Telescope} ({\it HST}), we searched for all detections of
LISM \ion{C}{2}$^{\ast}$ absorption.  We identify 13 sight lines with
23 individual \ion{C}{2}$^{\ast}$ absorption components, which provide
electron density measurements, the vast majority of which are new.  We
employ several strategies to determine more accurate \ion{C}{2} column
densities from the saturated \ion{C}{2} resonance line, including,
constraints of the line width from the optically thin
\ion{C}{2}$^{\ast}$ line, constraints from independent temperature
measurements of the LISM gas based on line widths of other ions, and
third, using measured \ion{S}{2} column densities as a proxy for
\ion{C}{2} column densities.  The distribution of electron densities
based on using \ion{S}{2} as a proxy for \ion{C}{2} is similar to the
distribution based on carbon alone, while significantly tighter, and
proves to be a promising technique to avoid grossly overestimating the
\ion{C}{2} column density based on the saturated line profile.  The
sample of electron densities appears consistent with a log-normal
distribution and an unweighted mean value of $n_e($\ion{C}{2}$_{\rm
SII}) = 0.11^{+0.10}_{-0.05}$ cm$^{-3}$.  Seven individual sight lines
probe the Local Interstellar Cloud (LIC), and all present a similar
value for the electron density, with a weighted mean of $n_e({\rm
LIC}) = 0.12 \pm 0.04$ cm$^{-3}$.  Two clouds, the NGP and Gem clouds,
show similar electron density properties as the LIC.  The Hyades
Cloud, a decelerated cloud at the leading edge of the platoon of LISM
clouds, has a significantly higher electron density than the LIC.
Observed toward G191-B2B, the high electron density may be caused by
the lack of shielding from such a strong radiation source.  No
evidence of a correlation between electron density and angular
separation of the sight line from the strongest extreme ultraviolet
radiation source, $\epsilon$ CMa, is found.  Given some simple
assumptions, the range of observed electron densities translates into
a range of thermal pressures, $P/k = 3300^{+5500}_{-1900}$
K~cm$^{-3}$.  This work greatly expands the number of electron density
measurements and provides important constraints on the ionization,
abundance, and evolutionary models of the local interstellar medium.

\end{abstract}


\section{Introduction}\label{intro}

More than a decade of use of high spectral resolution
($R\equiv\lambda/\Delta\lambda$~$\gtrsim$~50,000) ultraviolet (UV)
spectrographs with wide spectral coverage, such as the Goddard High
Resolution Spectrograph (GHRS) and particularly the Space Telescope
Imaging Spectrograph (STIS) onboard the {\it Hubble Space Telescope}
({\it HST}), has been a boon for observations of the local
interstellar medium (LISM).  Since the LISM, the collection of warm
gas in the immediate ($<$100 pc) vicinity of the Sun, is traversed by
all sight lines that extend beyond our solar system, the number of
observations with LISM absorption has grown tremendously.  The use of
this expanded database has been the impetus for many investigations
into the nature of the LISM, including the other papers in this
series, which present fits to LISM absorption in many ions (e.g.,
\ion{D}{1}, \ion{C}{2}, \ion{N}{1}, \ion{O}{1}, \ion{Mg}{2},
\ion{Al}{2}, \ion{Si}{2}, and \ion{Fe}{2},
\citealt{redfield02,redfield04sw}), temperature and turbulent velocity
measurements in LISM clouds \citep{redfield04tt}, and a dynamical
model of the LISM \citep{redfield07lism4}.

Due to the low column densities ($\log N($\ion{H}{1}$) \sim
16.8-18.3$) typical of local clouds, the ionization structure of
hydrogen is vital to understanding the physical structure and origins
of the LISM.  An accurate accounting of the ionized fraction of
hydrogen (and the ionization levels of all ions) is critical for
measuring abundances and the depletion of gas phase ions onto dust
grains.  Low column density clouds, (i.e., $\log N($\ion{H}{1}$) <
19.5$), like the LISM clouds, are not significantly shielded to
ionizing photons \citep{jenkins04,jenkins00,sofia98}.  The origin and
evolution of the Local Bubble, the $\sim$100 pc radius cavity in which
the warm LISM clouds reside \citep{lallement03}, is encoded in the
ionization structure.  Recent nondetections of high temperature lines
in the extreme UV \citep{hurwitz05} and the realization that soft
X-ray emission caused by the heliosphere \citep{lallement04} may
contribute to the emission formerly assigned to nearby hot gas
\citep{snowden90}, highlight the current challenges in understanding
the thermal structure and ionization level of the Local Bubble.  A
realistic Local Bubble can be modeled far from ionization equilibrium
\citep{breitschwerdt06} and the physical structure of the Local Bubble
can have a direct influence on the ionization structure of the warm
LISM clouds \citep{slavin02}.  Additionally, the structure of the
heliosphere, the interface between the LISM and the solar wind,
depends significantly on the ionization structure of the surrounding
interstellar medium \citep{muller06}.

The ionization fraction of hydrogen, $X({\rm H}) =
n($\ion{H}{2}$)/(n($\ion{H}{1}$) + n($\ion{H}{2}$))$, is computed
using the $n($\ion{H}{1}$) = n($\ion{He}{1}$) \times
N($\ion{H}{1}$)/N($\ion{He}{1}$)$ and $n($\ion{H}{2}$) \approx n_{\rm
e}$.  Typically, $n($\ion{He}{1}$)$ is derived from {\it in situ}
measurements of interstellar helium streaming through the heliospheric
interface into the inner solar system \citep{gloeckler04}, which of
course, is only a measurement of the helium density at one point in
the LISM.  $N($\ion{H}{1}$)/N($\ion{He}{1}$)$ is derived from extreme
UV observations of nearby white dwarfs (WDs), which contain the
ionization edge of \ion{He}{1}, and the continuum provides an estimate
of \ion{H}{1} \citep{dupuis95,barstow97}.  The weighted mean from this
work, based on nine white dwarf sight lines within 100\,pc, is
$N($\ion{H}{1}$)/N($\ion{He}{1}$)$ = $14.1 \pm 1.7$, although this is
a full sight line average, since this technique cannot separate
individual cloud components.
 
The remaining measurement in the calculation of the ionization
structure is the electron density, $n_e$.  The methods employed to
make this measurement utilize atomic transitions along interstellar
sight lines, which have the benefit of being able to resolve
individual absorbers if taken at high spectral resolution and can
provide a large number of measurements through various LISM
environments.  The ratio of magnesium ionization stage column
densities, $N($\ion{Mg}{2})$/N($\ion{Mg}{1}$)$, has provided a number
of $n_e$ measurements \citep[e.g.,][]{frisch90, lallement94, frisch94,
lallement97}.  However, this technique suffers from the requirement of
ionization equilibrium and a strong temperature dependence.
Alternatively, the ratio of the collisionally excited carbon line
column density to the resonance line column density,
$N($\ion{C}{2}$^{\ast})/N($\ion{C}{2}$)$, can provide $n_e$ estimates
without the need for ionization equilibrium and has a very weak
temperature dependence.  However, due to the weakness of the excited
\ion{C}{2}$^{\ast}$ absorption and saturation of the only available UV
\ion{C}{2} resonance line, few LISM sight lines have been analyzed
using this technique.  \citet{wood97} used high spectral resolution
($R \sim$ 100,000) spectra of $\alpha$ Aur to calculate $n_e$ along
the line of sight and demonstrated that the relatively simple
absorption profiles through the LISM provide an excellent opportunity
to make precise measurements of the electron density.
\citet{holberg99} used the same technique to measure the electron
density in ISM absorbers along the line of sight toward WD 1029+537, a
white dwarf $\sim$132\,pc away.  The excited absorption along other
LISM sight lines has been measured at moderate resolution ($R \sim$
20,000) with {\it Copernicus} \citep[e.g.,][]{york79} and the {\it Far
Ultraviolet Spectroscopic Explorer} ({\it FUSE}) \citep{lehner03},
although typically individual absorbers are not resolved, so the
estimates are full sight line averages.  A few $n_e$ measurements in
slightly warmer, and possibly local, \ion{O}{6}-bearing gas, derived
by comparisons of \ion{O}{6} emission \citep{dixon06} and absorption
\citep{lehner03}, have led to $n_e$ estimates of $\sim$0.2 cm$^{-3}$.

We present an inventory of high spectral resolution
\ion{C}{2}$^{\ast}$ detections along LISM sight lines from the
complete {\it HST} spectroscopic database.  We employ various
strategies to circumvent one of the challenges of using the
\ion{C}{2}$^{\ast}$/\ion{C}{2} ratio technique, namely obtaining a
reliable \ion{C}{2} column density from a saturated resonance line:
(a) Simultaneously fitting the optically thin \ion{C}{2}$^{\ast}$ and
saturated \ion{C}{2} profiles puts realistic constraints on the
Doppler width and column densities of the resonance line; (b) Use of
previously determined LISM temperatures and turbulent velocities,
typically derived from the same dataset \citep{redfield04tt},
constrains the Doppler width of the carbon lines; and (c) Use of
optically thin profiles of \ion{S}{2}, which has a similar ionization
potential as \ion{C}{2}, as a proxy of the \ion{C}{2} column density
\citep[e.g.,][]{oliveira03}.  This collection of new, high spectral
resolution UV observations of nearby stars presents an opportunity to
greatly expand the number of LISM electron density measurements and
probe the ionization structure of our most immediate interstellar
environment.

\section{Observations}\label{obs}

For our purpose, we are interested in sight lines toward stars within
100~pc that show interstellar absorption in both \ion{C}{2}
(1334.5323~\AA) and the \ion{C}{2}$^{\ast}$ doublet (1335.6627~\AA\
and 1335.7077~\AA).  We compiled all moderate to high resolution
observations of nearby stars with the {\it HST} spectrographs: GHRS
and STIS.  The complete sample includes 417 unique targets within
100~pc, almost half of which have spectra that cover the wavelength
region of \ion{C}{2} and \ion{C}{2}$^{\ast}$.  All these relevant
spectra were scrutinized for signs of \ion{C}{2}$^{\ast}$ absorption.
We found only 13 sight lines that show LISM absorption in both
transitions and list them in Table~\ref{starpar}.

The 100~pc distance limit is chosen to coincide with the approximate
extent of the Local Bubble \citep{lallement03}.  Observations of more
distant stars may be more difficult to analyze, as they are more
likely to traverse many absorbing clouds creating a blended line
profile.

Table~\ref{obspar} lists the observational parameters of the datasets
extracted from the {\it HST} Data Archive.  All observations are
necessarily moderate ($R \gtrsim$~20,000) to high ($R
\gtrsim$~100,000) spectral resolution in order to resolve narrow,
closely spaced interstellar absorption features and to increase the
likelihood of the detection of the weak \ion{C}{2}$^{\ast}$ line.

We analyzed LISM absorption in the \ion{S}{2} multiplet (1250.578 \AA,
1253.805 \AA, and 1259.518 \AA) in order to better constrain the
\ion{C}{2} column density.  Ten of the 13 targets have spectra that
include the \ion{S}{2} lines.  In addition, we fit the LISM absorption
in \ion{Mg}{1} (2852.9631 \AA) and the \ion{Mg}{2} doublet (2796.3543
\AA\ and 2803.5315 \AA) in order to further constrain the electron
density.  Only three of the 13 targets have spectra that include both
\ion{Mg}{1} and \ion{Mg}{2}.  Observations utilized for these analyses
are also included in Table~\ref{obspar}.

We reduced the GHRS data acquired from the {\it HST} Data Archive with
the CALHRS software package using the Image Reduction and Analysis
Facility \citep[IRAF;][]{tody93} and the Space Telescope Science Data
Analysis System (STSDAS).  We used the most recent reference
calibration files.  Many of the echelle observations were obtained in
the FP-SPLIT mode to reduce fixed-pattern noise.  The individual
readouts of the FP-SPLIT spectra are combined using a
cross-correlation procedure called HRS\_MERGE \citep{robinson92}.  The
reduction included assignment of wavelengths using calibration spectra
obtained during the course of the observations.  The calibration
spectra include either a WAVECAL, a direct Pt-Ne lamp spectrum used to
derive the dispersion relation, or a SPYBAL (Spectrum Y-Balance),
which only provides a zero-point wavelength offset.  Any significant
errors involved in the wavelength calibration are included in our
central velocity determinations.  The wavelength calibration of the
\ion{Fe}{2} spectrum of $\alpha$ Gru is used to calibrate the
wavelength solution of \ion{C}{2}, since no wavelength calibration of
this segment of the spectrum was taken at the time.

We reduced the STIS data acquired from the {\it HST} Data Archive
using the STIS team's CALSTIS software package written in IDL
\citep{lindler99}.  The reduction included assignment of wavelengths
using calibration spectra obtained during the course of the
observations.  We used the ECHELLE\_SCAT routine in the CALSTIS
software package to remove scattered light.  However, the scattered
light contribution is negligible in this spectral range, and does not
influence the uncertainties in our spectral analysis.

\section{Data Analysis and Line Profile Fitting}\label{lysis}

Figures~\ref{9plot} and \ref{4plot} show the \ion{C}{2} resonance and
the \ion{C}{2}$^{\ast}$ excited absorption lines observed toward the
13 targets listed in Table~\ref{starpar}.  The spectra (histogram) are
plotted in heliocentric velocity. Also plotted are the individual
interstellar component fits (dashed lines) and the total interstellar
absorption convolved with the instrumental profile (thick solid
lines).  As explained in Section~\ref{pro}, the resonance line plots
come from the simultaneous fits, and the excited line plots come from
the individual excited line fits.

The stellar continuum level (thin solid lines) illustrates our flux
estimates of the spectra without any interstellar absorption.  These
continua were determined by fitting polynomials to the spectral
regions adjacent to the observed interstellar features.  If the
intrinsic spectrum is a complex profile with intricate stellar
features, placing the continuum can be a difficult task.  All of our
target spectra have relatively smooth continua, and we are able to
easily reproduce the intrinsic continuum flux backgrounds with low
order polynomials.  Only two of the objects analyzed ($\alpha$ Aur and
EX Hya) include late-type stars, which are characterized by stellar
emission line spectra as opposed to smooth continuum spectra (e.g., as
seen in the white dwarf spectra), although their broad emission lines
allowed for simple continuum placement.

We used a Gaussian ISM absorption profile fitting algorithm to
determine the fits.  The program makes use of rest wavelengths and
oscillator strengths from \citet{morton03}, and the instrumental line
spread functions for GHRS and STIS spectra are taken from
\citet{gilliland94} and \citet{sahu99}, respectively.

\subsection{Fit Parameters}

Table~\ref{fitpar} lists the fit parameters measured, including
central velocity ($v$ [km s$^{-1}$]), Doppler width
($b$~[km~s$^{-1}$]) and the logarithm of the column density
($N$~[cm$^{-2}$]).  The central velocity is the mean radial velocity
of the absorbing component.  The Doppler width is a function of the
temperature ($T$~[K]) and turbulent velocity ($\xi$~[km s$^-1$]) of
the interstellar gas:
\begin{equation}
b^2=\frac{2kT}{m}+\xi^2=0.016629\frac{T}{A}+\xi^2,
\end{equation}
where $k$ is Boltzmann's constant, $m$ is the mass of the observed
ion, and $A$ is the atomic mass ($A_{\rm C} = 12.011$).  The column
density is a measure of the amount of material along the line of sight
to the target.

\subsection{Procedure}\label{pro}

For each target, we first fit the excited line alone.  Since the
excited line is actually a doublet, each paired absorption line
possesses the same intrinsic central velocity, Doppler width, and
column density; so these parameters are constrained accordingly to be
identical for both lines of the doublet.  We then fit the excited
(\ion{C}{2}$^{\ast}$) and resonance (\ion{C}{2}) lines simultaneously,
requiring the central velocities and Doppler widths to be identical.
Since the resonance line is typically saturated, $N$(\ion{C}{2}) is
not well constrained, but fitting the lines simultaneously allows us
to make use of information retained in the, albeit saturated,
resonance line, while retrieving vital constraints from the optically
thin excited line.  Thus, the final fit parameters for the resonance
line are derived from the simultaneous fits only.  For the excited
line fit parameters, we supplement the results from the simultaneous
fits with those from the individual fits in order to limit the impact
of systematic errors; these values are weighted means of the
parameters from both the simultaneous fits and individual excited line
fits.

In saturated lines, Doppler width and column density are tightly
coupled.  Thus, any available constraints on the $b$ value will yield
a more certain measurement for $N$, which in the case of \ion{C}{2}
and \ion{C}{2}$^{\ast}$, is critical in determining electron
densities.  Consequently, we use measured LISM temperatures and
turbulent velocities from \citet{redfield04tt} to constrain the
Doppler widths for the $\alpha$~Aur, $\eta$ UMa, and G191-B2B lines of
sight.  Unfortunately, LISM temperatures and turbulent velocities have
been determined for only a limited number of sight lines, so we are
able to include this type of constraint for only these three targets.

Figure~\ref{4plot} shows four white dwarfs (WD~0050-332, WD~0232+035,
WD~2309+105, and WD~1210+533) that each possess a resonance line
component that does not have an excited line counterpart.  In each
case, the Doppler width and column density for the lone resonance
component are largely determined by what is not accounted for by the
other paired components; the \ion{C}{2} measurements of the component
without a \ion{C}{2}$^{\ast}$ counterpart are poorly determined as
they lack the critical constraints provided by the optically thin
excited component.  Upper limits (3$\sigma$) are determined for the
\ion{C}{2}$^{\ast}$ column density for these components that show
absorption only in the resonance line.

The dominant source of systematic error, as mentioned earlier, is the
saturation of the \ion{C}{2} resonance line.  It is made evident upon
comparing our results for WD~0232+035 with those from
\citet{vennes00}.  Using the same datasets, \citet{vennes00} observed
two interstellar components toward WD~0232+035 with central velocities
of $3.1~\pm~0.2$~km~s$^{-1}$ and $17.6~\pm~0.9$~km~s$^{-1}$, in good
agreement with our measurements of $3.81~\pm~0.47$~km~s$^{-1}$ and
$17.4~\pm~4.6$~km~s$^{-1}$.  Their measured column density of
log$N$(\ion{C}{2}) = $14.32~\pm~0.57$ is a factor of 7 lower than our
measurement of log$N$(\ion{C}{2}) = $15.18~\pm~0.42$, but both have
large error bars, and the difference is just over 1$\sigma$.  In
contrast, the resonance line component that has no accompanying
excited line component is measured by \citet{vennes00} with a column
density log$N$(\ion{C}{2}) = $14.16~\pm~0.41$, a factor of 60 lower
than our measured value of log$N$(\ion{C}{2}) =
$15.96^{+0.32}_{-0.55}$, and a difference of almost 3$\sigma$.  The
critical factor in the disagreement concerns the treatment of the
Doppler width ($b$).  \citet{vennes00} allow for a range of $b$ values
(3.5--10.0~km~s$^{-1}$ and 4.0--10.0~km~s$^{-1}$ for the
$\sim$3~km~s$^{-1}$ and $\sim$17~km~s$^{-1}$ components,
respectively), while we allow $b$ to remain an independent variable.
Such a wide range of $b$ values appears to be unnecessary; for our
\ion{C}{2} fits to all sight lines, the weighted mean Doppler width is
$b=4.30~\pm~0.87$~km~s$^{-1}$, and inconsistent with a large b value
for WD0232+035.  In addition, our fit to the \ion{S}{2} lines, which
also shows two components at similar velocities (4.11 $\pm$
0.64~km~s$^{-1}$ and 16.8 $\pm$ 3.8~km~s$^{-1}$), has a weighted mean
Doppler width of $b=3.95~\pm~0.79$~km~s$^{-1}$, also inconsistent with
a large Doppler width.

Simultaneously fitting the unsaturated excited line with the
accompanying saturated resonance line is the critical property of our
method, which allows us to more tightly constrain the observed column
density of the saturated line than if we were fitting it alone.
Essentially, the optically thin excited line provides the constraint
on the Doppler width, which thereby constrains the acceptable range of
column densities.

\subsection{Component Determination}

In determining the number of absorption components, we use the fewest
number of components that produce a satisfactory fit.  In most cases,
the resolution is high enough and the differences between component
velocities are great enough such that discernment of components is
almost trivial.  Other times, however, it is not so easily determined.
Components are added if a clear asymmetry is detected in either the
\ion{C}{2} and \ion{C}{2}$^{\ast}$ profiles (e.g., $\eta$ UMa,
WD~1210+533, IX Vel), or if other ions indicate that additional
components are required (e.g., $\alpha$~Gru, $\rho$ Lup).  Our general
attitude toward this part of the analysis is to approach the
absorption feature with Ockham's razor in hand, so as to cut away any
unjustified components.

\subsection{\ion{S}{2}}

We are motivated to find an alternative means of estimating the
\ion{C}{2} column density, due to the difficulty of obtaining it
directly from the strongly saturated resonance line.  We use the
optically thin \ion{S}{2} triplet, located near the \ion{C}{2} lines,
as a proxy for \ion{C}{2} due to their similarity in ionization
potential, which is 24.4 eV for \ion{C}{2} and 23.3 eV for \ion{S}{2},
and their similarity in the ratio of their ionization and
recombination rates (defined as $P$ in \citet{sofia98}, where
\ion{S}{2} and \ion{C}{2}, among other ions, are compared in their
Figure~3).  Ten of the 13 sight lines with \ion{C}{2}$^{\ast}$
absorption also have spectra that cover the \ion{S}{2} wavelength
range.  We fit all three \ion{S}{2} lines simultaneously, and due to
the range of opacities of the three optically thin lines, we are able
to measure an accurate \ion{S}{2} column density.  The spectra and
fits are shown in Figures~\ref{fig:9plots2} and \ref{fig:4plots2}, and
the fit parameters are listed in Table~\ref{tab:s2fitpar}.

In order to convert from $N($\ion{S}{2}$)$ to $N($\ion{C}{2}$)$, we
need to take into account the different abundances and depletion
levels of sulfur and carbon in the LISM:
\begin{equation}
N({\rm CII}_{\rm SII})=N({\rm
SII})\times 10^{[{\rm C}_\odot + D({\rm C})] - [{\rm S}_\odot + D({\rm S})]},
\label{eq:s2toc2}
\end{equation}
where $N$(\ion{C}{2}$_{\rm SII}$) is the estimated \ion{C}{2} column
density based on \ion{S}{2} as a proxy, $N$(\ion{S}{2}) is our
measured column density of \ion{S}{2}, $C_{\odot} = 8.39 \pm 0.05$ and
$S_{\odot} = 7.14 \pm 0.05$ are the solar abundances of carbon and
sulfur \citep{asplund05}, and $D($C$)$ and $D($S$)$ are the depletion
levels \citep{jenkins04}.  Here $D($S$) \sim 0.0$, as has been
detected in local and more distant ISM \citep{lehner03,welty99}, and
$D($C$) \sim -0.17 \pm 0.19$ from \citet{jenkins04}, where the error
incorporates some of the natural variation in the ISM (e.g., $D($C$)
\sim 0.0$ \citealt{cardelli96}; $\sim$--0.2 \citealt{lehner03};
$\sim$--0.4 \citealt{welty99}).

\begin{figure}
\figurenum{1a}
\epsscale{.7}
\plotone{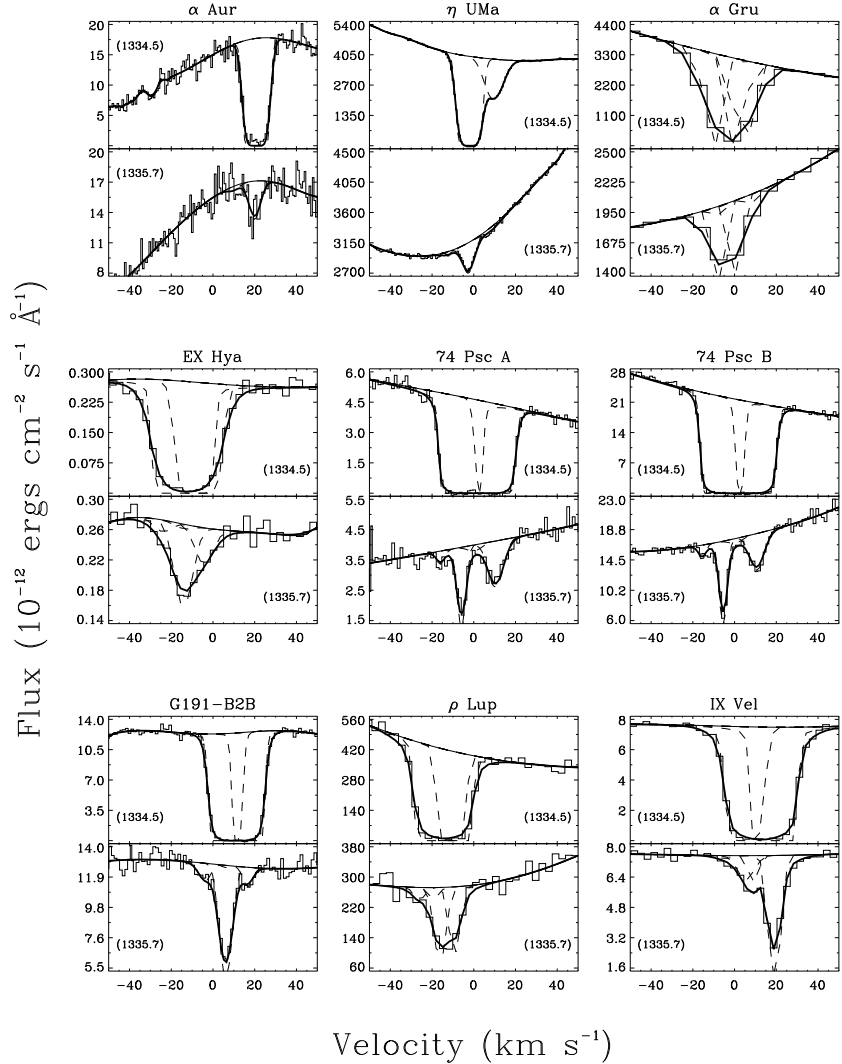}
\caption{Fits of the interstellar \ion{C}{2} (1334.5~\AA) and
\ion{C}{2}$^{\ast}$ (1335.6627~\AA\ and 1335.7077~\AA) absorption
toward 13 nearby stars.  The ratio of the column densities are used to
estimate the electron density for each component.  The name of the
target star is given above each group of plots, and the wavelength (in
Angstroms) of each line is provided within each plot.  Both the
1335.6627~\AA\ and 1335.7077~\AA\ \ion{C}{2}$^{\ast}$ lines are shown
in the bottom plot.  Although the 1335.6627~\AA\ line is weak, it is
evident as the blueward component in some spectra (e.g., G191-B2B, 74
Psc A and B, and $\rho$ Lup).  The data are shown in histogram form.
The thin solid lines are our estimates of the intrinsic stellar flux
across the absorption feature.  The dashed lines are the best-fit
individual absorption lines before convolution with the instrumental
profile.  The thick solid line represents the combined absorption fit
after convolution with the instrumental profile.  The spectra are
plotted versus heliocentric velocity.  The parameters for these fits
are listed in Table~\ref{fitpar}.\label{9plot}}
\end{figure}

\begin{figure}
\epsscale{.7}
\figurenum{1b}
\plotone{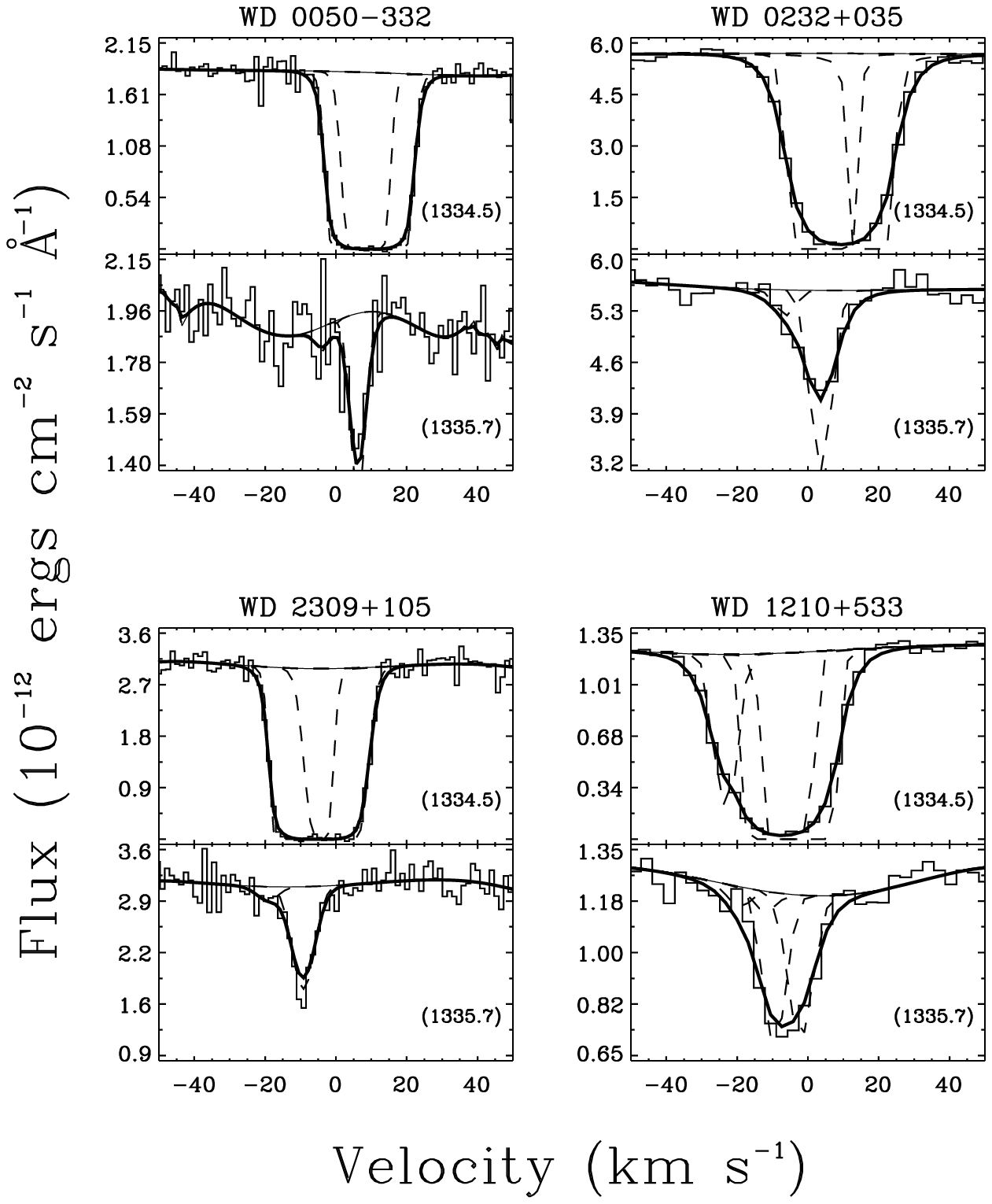}
\caption{\label{4plot}}
\end{figure}

\begin{figure}
\epsscale{.7}
\figurenum{2a}
\plotone{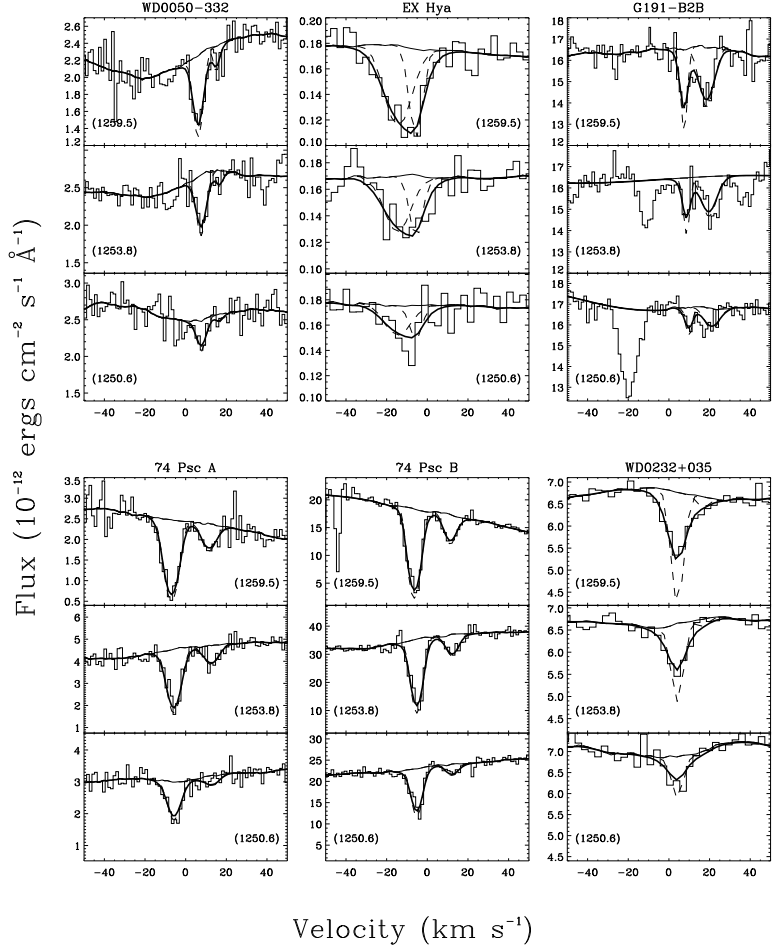}
\caption{Similar to Figure~\ref{9plot}, but for fits to interstellar
\ion{S}{2} absorption toward 10 of the 13 targets which show
\ion{C}{2}$^{\ast}$ absorption.  The column density of the optically
thin \ion{S}{2} profiles are used to estimate the column density of
\ion{C}{2} along the line of sight.  All three \ion{S}{2} lines are
fit simultaneously.  The parameters for these fits are listed in
Table~\ref{tab:s2fitpar}.
\label{fig:9plots2}}
\end{figure}

\begin{figure}
\figurenum{2b}
\epsscale{.7}
\plotone{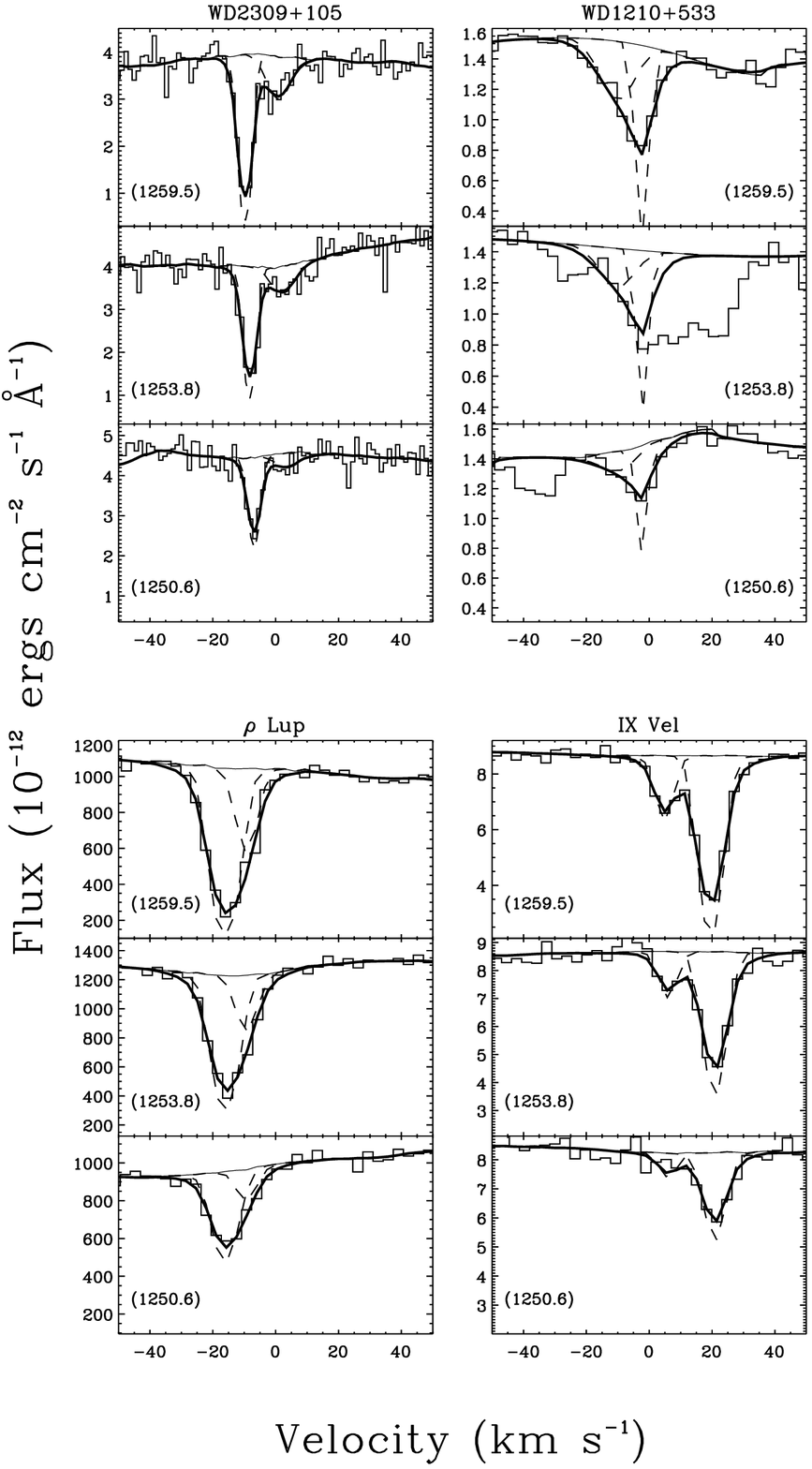}
\caption{\label{fig:4plots2}}
\end{figure}

The \ion{C}{2} column density estimated using this technique is listed
in Table~\ref{fitpar}, and the comparison with the direct \ion{C}{2}
column density based on the \ion{C}{2} and \ion{C}{2}$^{\ast}$ fits
shown in Figures~\ref{9plot} and \ref{4plot} is exhibited in
Figure~\ref{fig:cscomp}.  We present two errors based on this
calculation: the first simply propagates the error in $N({\rm SII})$,
while the second includes the errors in the solar abundances
\citep{asplund05} and the natural variation of depletions
\citep{jenkins04}.  Figure~\ref{fig:cscomp} demonstrates that
\ion{S}{2} is a reasonable proxy for \ion{C}{2}, where 12/20 (60\%)
agree within 2$\sigma$, and those that disagree by $>$2$\sigma$ are
all overestimated by measuring the strongly saturated \ion{C}{2}
resonance line directly.  Forcing $N({\rm CII}) = N({\rm CII}_{\rm
SII})$ in fitting \ion{C}{2} and \ion{C}{2}$^{\ast}$ produces viable
fits to the data that are indistinguishable from those shown in
Figures~\ref{9plot} and \ref{4plot}.  The discrepancy between the
$N({\rm CII})$ and $N({\rm CII}_{\rm SII})$ originates from the
well-known problem when on the flat part of the curve-of-growth, that
very small changes in the Doppler width can produce very large changes
in the column density.  Despite the complication of using the ISM
average depletions for carbon and sulfur, estimating the \ion{C}{2}
column density is likely more accurately achieved by using \ion{S}{2}.

\subsection{\ion{Mg}{1} and \ion{Mg}{2}}

Spectra of three ($\eta$ UMa, $\alpha$ Gru, and G191-B2B) of the 13
sight lines with \ion{C}{2}$^{\ast}$ absorption also contain the
\ion{Mg}{1} and \ion{Mg}{2} lines.  We fit the optically thin
\ion{Mg}{1} line separately and the two marginally saturated
\ion{Mg}{2} lines simultaneously.  The spectra and fits are shown in
Fig~\ref{fig:mgplot} and the fit parameters are listed in
Table~\ref{tab:mgfitpar}.  Due to the strength of the \ion{Mg}{2}
lines and relative weakness of the \ion{Mg}{1}, not all \ion{Mg}{2}
components are detected in \ion{Mg}{1}.  Although those that are, have
consistent velocities and Doppler widths, indicating both ions are
likely part of the same collections of gas.

\subsection{Individual Sight Lines}

{\it $\alpha$~Aur.}---The line of sight toward $\alpha$~Aur is well
known to exhibit a strong single interstellar absorption component
\citep{linsky93,linsky95} that has been identified with the Local
Interstellar Cloud (LIC; \citealt{lallement95}).  These
characteristics, complemented by the star's proximity, contributed to
it being the first LISM sight line with a $n_e$ measurement based on
absorption detected in \ion{C}{2} and \ion{C}{2}$^{\ast}$
\citep{wood97}.  The first fit plotted in Figure~\ref{9plot} is for
$\alpha$~Aur.  As noted by \citet{wood97}, who analyzed the same data,
the low signal-to-noise and the saturation of the resonance line make
the measurement of a precise Doppler width difficult.  For this
reason, we use the independently determined LISM temperature and
turbulent velocity along this line of sight to force the Doppler width
to be $3.48^{+0.15}_{-0.19}$ km~s$^{-1}$ \citep{redfield04tt}.  Our
measured central velocity of $v=20.78\pm0.28$ km~s$^{-1}$ agrees with
the $20.2\pm1.5$ km~s$^{-1}$ measured by \citet{wood97}.  Also, we
measure column densities of log$N$(\ion{C}{2})$=14.67^{+0.14}_{-0.22}$
and log$N$(\ion{C}{2}$^{\ast}$)$=12.62\pm0.07$, which agree with their
log$N$(\ion{C}{2})$=14.8\pm0.3$ and
log$N$(\ion{C}{2}$^{\ast}$)$=12.64\pm0.07$, respectively.

{\it $\eta$~UMa.}---The $\eta$~UMa sight line shows two components.
The Doppler width of the absorption observed toward this target is
constrained based on independent temperature and turbulent velocity
information \citep{redfield04tt}.  For the first component, $b$ is
fixed at $3.76^{+0.09}_{-0.11}$ km~s$^{-1}$, and for the second
component, $b$ is fixed at $5.60^{+0.09}_{-0.11}$ km~s$^{-1}$.  All
\ion{C}{2} fit parameters from \citet{redfield04sw} agree with our
measurements.  Both components are detected in \ion{Mg}{1} and
\ion{Mg}{2}, although the redward component is very weak and has a
poorly determined central velocity and Doppler width.  Indeed, the
blueward ($\sim$--2 km~s$^{-1}$) component is consistent in velocity
for both carbon and magnesium, but the measured velocity of the
redward component is significantly discrepant between the two,
although the errors are large.  \citet{frisch06aas} provided an
in-depth analysis of this particular sight line, including a
measurement of $n_e \sim 0.1$ cm$^{-3}$.

{\it $\alpha$~Gru.}---The spectra of $\alpha$~Gru were taken at the
lowest resolution allowed in our survey.  Based on this observation
alone, it is difficult to identify the appropriate number of
absorption components, but on the merit of past high spectral
resolution observations of other ions (e.g., \ion{Mg}{2},
\ion{Fe}{2}), three interstellar components have been determined along
this line of sight \citep{redfield02}.  We fit three components
accordingly and use the same difference in central velocities between
the three components as in \citet{redfield02} to constrain our
\ion{C}{2} fit.  We use the wavelength solution of \ion{Fe}{2} in
order to calibrate the \ion{C}{2} spectrum, which had no wavelength
calibration image at the same wavelength.  The errors are large for
Doppler width and column density due to the blending of the three
components.  Our measurements are consistent, though more reliable,
than those by \citet{redfield04sw}, in which the \ion{C}{2} resonance
line was fit alone.  Our \ion{Mg}{2} measurements agree well with
those of \citet{redfield02}.

{\it WD~0050-332.}---Using low-resolution ($R \sim $10,000; $\Delta$$v
\sim 30$ km~s$^{-1}$) International Ultraviolet Explorer ({\it IUE})
observations, \citet{holberg98} find interstellar \ion{C}{2} toward
WD~0050-332 at $10.69\pm3.20$ km~s$^{-1}$.  At higher resolution, our
\ion{C}{2} and \ion{S}{2} observations resolve this feature into two
components at $\sim$6.3 and $\sim$12.6 km~s$^{-1}$.  In agreement with
our claim of two components, two components have also been observed in
\ion{N}{1} and \ion{Si}{2} by \citet{oliveira05} at $\sim$6.8 and
$\sim$16.8 km~s$^{-1}$.  In \ion{C}{2}$^{\ast}$, however, we observe
only the stronger $\sim$6.3 km~s$^{-1}$ component.  Tentative
detections of highly ionized components associated with a
circumstellar shell have been observed toward this target
\citep{holberg98,bannister03}.

{\it EX~Hya and IX~Vel.}---Both targets have been scrutinized for
observational evidence for circumbinary disks \citep{belle04}.  In
this search, which used the same datasets employed here, they identify
absorption in \ion{C}{2}, \ion{C}{2}$^{\ast}$, and \ion{S}{2},
although the absorption was not identified with circumbinary
absorption for either star.  Instead, \citet{belle04} attributed the
observed absorption to the LISM and fit the profile with a single
Gaussian function, although particularly in the case of IX Vel, two
components are clearly present.  We fit both EX Hya and IX Vel with
two components, motivated by clear asymmetries in the
\ion{C}{2}$^{\ast}$ and \ion{S}{2} absorption.  The column-weighted
velocity of our components match fairly well with their estimates for
EX Hya.  However, the weaker \ion{C}{2}$^{\ast}$ transition
(1335.6627~\AA) appears to be used as the wavelength standard instead
of the stronger transition (1335.7077~\AA), which overestimates the
velocity of absorption by $\sim$10~km~s$^{-1}$.  This correction
improves the agreement of their \ion{C}{2}$^{\ast}$ velocity with the
other lines they measured and with our measurements.  Their IX Vel
estimates are significantly different from ours and are likely due to
the poor approximation of a single Gaussian to the absorption profile.
\citet{linnell07} also note the presence of interstellar absorption in
\ion{C}{2} and \ion{S}{2} in the spectrum of IX~Vel.

{\it G191-B2B.}---This is the third target which the Doppler width can
be constrained from independent measurements of temperature and
turbulent velocity \citep{redfield04tt}.  For the first and second
components, $b=4.10^{+0.29}_{-0.31}$ km~s$^{-1}$ and
$b=3.43^{+0.21}_{-0.26}$ km~s$^{-1}$, respectively.  The column
density measured by \citet{redfield04sw} for the $\sim$17~km~s$^{-1}$
component, log$N$(\ion{C}{2})$=15.70\pm0.36$, is consistent with our
measurement, log$N$(\ion{C}{2})$=15.42\pm0.17$.  The
$\sim$6~km~s$^{-1}$ component is in significant disagreement, with
their estimate at log$N$(\ion{C}{2})$=16.12\pm0.33$ and ours at
log$N$(\ion{C}{2})$=14.8^{+0.3}_{-1.2}$.  The measurements by
\citet{redfield04sw} were obtained from fitting the saturated
\ion{C}{2} resonance line only, which typically overestimates the
column density.  The current fits are more reliable since both
\ion{C}{2} and \ion{C}{2}$^{\ast}$ profiles are used simultaneously,
which more tightly constrain the fit parameters.  We measure
log$N$(\ion{C}{2}$^{\ast}$)$=12.28\pm0.09$ for the second component,
which matches a velocity believed to be attributed to the LIC
\citep{lallement95} in this direction.  Lying angularly close
($\sim$2.5$^{\circ}$) and believed to traverse the LIC is the line of
sight towards $\alpha$~Aur.  We measure its column density to be
slightly greater with
log$N$(\ion{C}{2}$^{\ast}$)$=12.62\pm0.07$. Since the distance to
$\alpha$~Aur is 20\% that to G191-B2B, it is clear that the LIC does
not extend beyond $\sim$13 pc in that direction.  \ion{Mg}{1},
\ion{Mg}{2}, and \ion{S}{2} are also measured along this line of sight
using two components.  The velocities agree between magnesium and
sulfur, although they are slightly larger than those measured with
carbon.  Indeed, these \ion{C}{2} and \ion{C}{2}$^{\ast}$ fits differ
in velocity even from the \ion{C}{2} fits alone from
\citet{redfield04sw}.  Again, since the current measurements are also
constrained by the optically thin excited line, these measurements are
more reliable.  Why the G191-B2B absolute radial velocities disagree
among different ions is unknown, but may be explained by systematic
wavelength calibration issues, as the difference in radial velocity
between the components agrees for all ions, and the spectra used for
different ions are coadditions of different individual observations at
various grating settings.  \citet{lemoine96} also measure LISM
absorption for two dominant components at $\sim$9.9 and $\sim$20.6
km~s$^{-1}$ and derive roughly similar column densities for
\ion{Mg}{2} and \ion{C}{2}.

{\it WD~0232+035.}---\citet{dupree82} identified interstellar
absorption in \ion{O}{1}, \ion{Si}{2}, and \ion{N}{1} while suggesting
the presence of a Str\"omgren sphere due to absorption observed in
\ion{C}{4}.  There have also been continued discussions of high
ionization circumstellar features for this star
\citep[e.g.,][]{vennes94,holberg98,bannister03}.  \citet{holberg98} observe
LISM absorption of \ion{C}{2} at $v=-4.28\pm2.78$ km~s$^{-1}$ with low
resolution {\it IUE} spectra.  Using the same datasets as
\citet{vennes00}, we resolve two components in both \ion{C}{2} and
\ion{S}{2} at $\sim$3.9 and $\sim$17.0 km~s$^{-1}$, which match their
$\sim$3.1 and $\sim$17.6 km~s$^{-1}$ components very well.  We find
evidence of the $\sim$17.0 km~s$^{-1}$ component in \ion{S}{2}, most
notably in the strong 1259.5~\AA\ line, whereas \citet{vennes00} only
fit the $\sim$3.9 km~s$^{-1}$ component.  We observe, just as they do,
only the $\sim$3.9 km~s$^{-1}$ component in the \ion{C}{2}$^{\ast}$
excited line.

{\it WD~2309+105.}---\citet{holberg98} use low-resolution {\it IUE}
spectra to identify one LISM component in \ion{C}{2}$^{\ast}$ at
$v=-8.27\pm3.19$ km~s$^{-1}$ and in all three \ion{S}{2} lines at
velocities ranging from --14.5 to --6.7 km~s$^{-1}$.  Two components
are seen in high resolution {\it HST} spectra analyzed by
\citet{oliveira03} in \ion{N}{1}, \ion{S}{2}, and \ion{Si}{2}, while
only the redward component is detected in \ion{C}{2}$^{\ast}$.  Using
the same dataset, we concur, and derive nearly identical fit
parameters.  Our central velocities are slightly different; ours are
measured at $\sim$--8.5 $\pm$ 1.2 and $\sim$1.3 $\pm$ 1.8 km~s$^{-1}$
and theirs at $\sim$--9.8 and $\sim$--0.6 km~s$^{-1}$, but the
differences are minimal given the large errors and almost identical
differential velocity.  Only the $\sim$--8.5 km~s$^{-1}$ component is
detected in \ion{C}{2}$^{\ast}$ in agreement with the {\it IUE}
detection by \citet{holberg98}.

{\it WD~1210+533.}---Although only one component could be discerned in
the low resolution {\it IUE} spectra, \citet{holberg98} find
\ion{C}{2} absorption at $v=-7.32~\pm~3.21$ km~s$^{-1}$ and
\ion{C}{2}$^{\ast}$ absorption at $v=-12.14~\pm~5.92$ km~s$^{-1}$.  We
detect three components in \ion{C}{2}, while only the two redward
components are seen in \ion{S}{2}, at velocities of $\sim$--23.4,
$\sim$--9.3, and $\sim$--3.0 km~s$^{-1}$.  The \ion{S}{2} 1253.8~\AA\
line suffers from considerable contamination, although the two
components are well characterized in the other two \ion{S}{2} lines.

{\it $\rho$~Lup.}---The two components characterizing the excited line
of $\rho$~Lup are similar in Doppler width and in column density,
lending difficulty to distinguishing the two despite a $\sim$7
km~s$^{-1}$ separation.  Fortunately, the asymmetric \ion{S}{2}
absorption profile allows us to discern two LISM components.  We use
velocities determined from these measurements to constrain our
\ion{C}{2} fits.  \citet{welsh05} observe two components in
\ion{Na}{1} (5890.0~\AA) and three components in \ion{Fe}{2}
(1608.5~\AA), \ion{S}{2} (1253,1259~\AA), and \ion{Al}{2}
(1670.8~\AA).  Although we only find evidence for two components in
\ion{C}{2} and \ion{S}{2}, our components agree fairly well, with
velocities $\sim$--16.1 and $\sim$--9.1 km~s$^{-1}$.  \citet{welsh05}
present a thorough discussion of cloud distances and identify the
$\sim$--9.1 km~s$^{-1}$ absorption with material at $\sim$90 pc while
identifying the $\sim$--16.1 km~s$^{-1}$ absorption with material at
the neutral boundary of the Local Bubble in the direction toward
$\rho$ Lup.

\section{Electron Density}

\subsection{Estimation Based On \ion{C}{2}$^{\ast}$}

Table~\ref{electron} lists the measured electron densities, $n_e$,
along the lines of sight towards the 13 targets analyzed.  The values
listed in column four of this table are calculated using a method
which compares the column densities of the resonance and excited lines
of \ion{C}{2}.  Our use of this method is similar to that implemented
by \citet{spitzer93} and \citet{oliveira03} but most parallels that of
\citet{wood97}.

The \ion{C}{2} resonance absorption line at 1334.5323~\AA\ corresponds
to the transition from the ground state ($J=1/2$), while the
\ion{C}{2}$^{\ast}$ excited absorption lines at 1335.6627~\AA\ and
1335.7077~\AA\ correspond to the transition from the excited state of
the fine-structure doublet ($J=3/2$).  Collisions with electrons are
responsible for populating the excited state, and hence the ratio of
the column densities of the two lines is proportional to the electron
density.  For a detailed discussion of how the fine structure of
absorption lines can be used to determine density, see
\citet{bahcall68}.

The following relation is derived from thermal equilibrium between
collisional excitation of the $J=3/2$ state and radiative
de-excitation:
\begin{equation}
\frac{N({\rm CII^{\ast}})}{N({\rm CII})}=\frac{n_eC_{12}(T)}{A_{21}}.
\end{equation}
The effect of collisional de-excitation at these densities and
temperatures is negligible and therefore not included in the equation
above.  $N$(\ion{C}{2}) and $N$(\ion{C}{2}$^{\ast}$) are the column
densities of the resonance and excited lines, respectively.  The
calculation of the electron density using \ion{S}{2} as a proxy for
\ion{C}{2}, simply replaces $N$(\ion{C}{2}) with $N($\ion{C}{2}$_{\rm
SII})$, as derived from Equation~\ref{eq:s2toc2}.  The electron
densities based on $N($\ion{C}{2}$_{\rm SII})$ are also listed in
Table~\ref{electron} (column five).  The radiative de-excitation rate
coefficient is $A_{21}=2.29\times10^{-6}$~s$^{-1}$, as listed by
\citet{nussbaumer81}.  The collision rate coefficient can be expressed
is cgs units as
\begin{equation}
C_{12}(T)=\frac{8.63\times10^{-6}\Omega_{12}}{g_1T^{0.5}}\exp{\Biggl(-\frac{E_{12}}{kT}\Biggr)},
\end{equation}
where the statistical weight of the ground state $g_1=2$, and the
energy of the transition $E_{12}=1.31\times10^{-14}$~ergs. Since the
collision strength $\Omega_{12}$ has a very weak temperature
dependence, for all targets we let $\Omega_{12}=2.81$ \citep{hayes84},
calculated at a temperature of 7000~K, very similar to the LISM
average of 6680~K \citep{redfield04tt}.

\subsection{Estimation Based On \ion{Mg}{1} and \ion{Mg}{2}}

Three of our targets have both \ion{Mg}{1} and \ion{Mg}{2} LISM
absorption measurements (see Table~\ref{tab:mgfitpar}).  We estimate
the electron density by following the procedure detailed by
\citet{lallement97} and \citet{frisch94}.  Assuming equilibrium, the
electron density can be estimated by formulating the balance of
ionization and recombination between neutral and singly ionized
magnesium:
\begin{equation}
\frac{N({\rm MgII})}{N({\rm MgI})}=\frac{\Gamma + n_{\rm e}\sigma_{\rm ex}}{n_{\rm e}\alpha},
\end{equation}
where $\Gamma$ is the photoionization rate of \ion{Mg}{1},
$\sigma_{\rm ex}$ is the formation rate of \ion{Mg}{2} based on charge
exchange, $\alpha$ is the total recombination rate, and we have
assumed that $N($\ion{Mg}{2}$)/N($\ion{Mg}{1}$) =
n($\ion{Mg}{2}$)/n($\ion{Mg}{1}$)$ and $n($\ion{H}{2}$) = n_{\rm e}$.
We use a photoionization rate of $\Gamma = 6.1 \times
10^{11}$~s$^{-1}$ \citep{sofia98,jenkins00bcma}.  Although
photoionization typically dominates over charge exchange, at high
enough densities, the charge exchange contribution can be significant.
We have used the relationship derived by \citet{allan88}, $\sigma_{\rm
ex} = 1.74 \times 10^{-9} \exp(-22100/T)$ cm$^3$~s$^{-1}$.  The total
recombination rate, similar to \citet{frisch94}, is the sum of the
radiative recombination rate from \citet{aldrovandi73} and dielectric
recombination rate from \citet{nussbaumer86} and \citet{mazzotta98}.
All recombination rates are strongly dependent on temperature.
Table~\ref{electron} lists the calculated electron densities based on
the ratio of \ion{Mg}{1} and \ion{Mg}{2} (column six).

\subsection{Previous Measurements}

Electron density estimates have been made by other researchers for a
few of the sight lines.  Toward $\alpha$~Aur, our electron density
($n_e=0.140^{+0.060}_{-0.059}$ cm$^{-3}$) agrees well with the
measured $n_e=0.11^{+0.12}_{-0.06}$ cm$^{-3}$ by \citet{wood97}.  An
interstellar temperature of 6700~K \citep{redfield04tt} is included in
our determination of the electron density.  \citet{vennes00} obtained
an electron density estimate toward WD~0232+035 by assuming a wide
range of possible $b$ values which resulted in a density of $n_e=0.36$
cm$^{-3}$ for the interstellar component near $3.8$ km~s$^{-1}$.  This
is slightly higher, but within 2$\sigma$ of our estimates,
$n_e=0.10^{+0.17}_{-0.6}$ cm$^{-3}$ and $n_e({\rm CII}_{\rm
SII})=0.21^{+0.14}_{-0.08}$ cm$^{-3}$.  \citet{oliveira03} provide an
estimate of $n_e = 0.1 \pm 0.01$ cm$^{-3}$ for the --9.7 km~s$^{-1}$
component toward WD~2309+105, based on using \ion{S}{2} as a proxy of
\ion{C}{2} and assuming a temperature of 8000~K.  Our measurement
using the same technique, assuming the LISM average temperature of
6680~K, matches quite well, $n_e = 0.084^{+0.052}_{-0.033}$ cm$^{-3}$.

\section{Discussion}\label{disc}

Figures~\ref{fig:cne}--\ref{hist} summarize the results of our
electron density measurements.  Figure~\ref{fig:cne} shows the
measured electron density as a function of the \ion{C}{2}$^{\ast}$
column density.  Electron density estimates using both the saturated
\ion{C}{2} resonance line and \ion{S}{2} as a proxy for \ion{C}{2} are
shown.  The measurements using \ion{S}{2} are more precise, even when
we include systematic errors of the variation of depletion, than the
\ion{C}{2} measurements since \ion{S}{2} is optically thin.
Figure~\ref{den2} displays the same distribution of data points,
except instead of electron density, we plot the resonance line column
density versus the excited line column density.  Overplotted are
constant-density contours for the mean temperature of the LISM.  Both
plots show that the LISM electron density measurements are relatively
tightly log-normal distributed about the unweighted mean value of
$\sim$0.1~cm$^{-3}$.  Figure~\ref{hist} further emphasizes the tight
distribution of electron density measurements about the mean.  In
particular, by using optically thin \ion{S}{2} lines as a proxy for
\ion{C}{2}, we avoid systematically high column density measurements
commonly derived from saturated line profiles, which ultimately lead
to very low electron density determinations.  For this reason, in the
electron density distribution derived from \ion{S}{2}, we lose the low
electron density tail evident with the \ion{C}{2} measurements, and
the LISM average tightens to an unweighted mean of
0.11$^{+0.10}_{-0.05}$~cm$^{-3}$, where the 1$\sigma$ errors are the
dispersion about the mean value.  All electron densities derived 
using \ion{S}{2} as a proxy range from 0.07--0.80~cm$^{-3}$.  The
distribution of $n_e($\ion{C}{2}$_{\rm SII})$ measurements matches well
the $n_e($\ion{C}{2}$)$ distribution, but it is significantly tighter,
since the gross overestimates of the \ion{C}{2} column density are
avoided.  This is a strong endorsement for using \ion{S}{2} as a proxy
for \ion{C}{2}, in order to make more accurate electron density
measurements.

Figure~\ref{fig:comp} shows the solutions of electron density as a
function of temperature for the \ion{C}{2}$^{\ast}$/\ion{C}{2}
technique, which does not require ionization equilibrium and has a
very modest temperature dependence, and for the
\ion{Mg}{1}/\ion{Mg}{2} technique, which does require ionization
equilibrium and has a very strong temperature dependence.  This type
of comparison was used by \citet{gry01} to measure the properties
along the sight line toward $\epsilon$ CMa.  Also shown in
Figure~\ref{fig:comp}, for several of the sight lines, is an
independent measurement of the temperature of the gas based on the
widths of absorption lines of different atomic mass
\citep{redfield04tt}.  Ideally, all three measurements should converge
at the same temperature and provide a single estimate of the electron
density.

\subsection{LISM Cloud Properties}

The last column of Table~\ref{electron} indicates the interstellar
cloud(s) by which the observed components may be identified.  The
criteria for identification include agreement with the predicted
projected velocity and spatial distribution of a given LISM cloud
\citep{redfield07lism4}.  Components for some of our targets
($\alpha$~Aur, $\alpha$~Gru, and G191-B2B for the LIC; $\eta$~UMa for
the NGP cloud; G191-B2B for the Hyades cloud) had been previously
identified and assigned to specific clouds by \citet{redfield07lism4}.
Many of the remaining components could also be identified with known
LISM clouds (i.e., the observed radial velocity of absorption is
within 3$\sigma$ of the predicted projected velocity, and the line of
sight traverses the spatial distribution of the cloud).  Those clouds
listed in parentheses also agree in velocity, but the sight line only
passes near (within 20 degrees of) the given cloud boundary.  Given
the similarities of the LISM cloud velocities, it is not always easy
to uniquely identify cloud membership, although of the 23
\ion{C}{2}$^{\ast}$ components, 11 can be firmly identified with a
specific LISM cloud.

The Local Interstellar Cloud (LIC) is the collection of gas that
dominates LISM absorption line observations because it is detected in
a large fraction of the sky.  The Sun is currently located just
outside of the LIC \citep{redfield00,redfield07lism4}.  While LIC
material has likely surrounded the Sun for the last $\sim$100,000
years, given the relative motion and location of the LIC and the Sun,
the solar system has moved, or will very shortly move, into a different
interstellar environment.  However, the proximity of the LIC means
that many LISM sight lines will probe this material.  Indeed, 7 of the
11 identified cloud members are of the LIC.  The LIC $n_e$
measurements are shown by the shaded histogram in Figure~\ref{hist},
where all sight lines give a similar measurement.  The
$n_e($\ion{C}{2}$_{\rm SII})$ measurements are used to calculate the
mean when available, while $n_e($\ion{C}{2}$)$ is used for
$\alpha$~Aur and $n_e($\ion{Mg}{2}$/$\ion{Mg}{1}$)$ is used for
$\alpha$~Gru.  The weighted mean is $n_e({\rm LIC}) = 0.12 \pm 0.04$
cm$^{-3}$.  Dramatic variation in electron density within the LIC is
not observed.  However, more sight lines would be required to do a
more involved investigation of intracloud variability.

Other electron density components are assigned to various other
clouds, but none have as many measurements as the LIC.  The North
Galactic Pole (NGP) Cloud has two $n_e$ measurements which agree quite
will with each other at $\sim$0.09 cm$^{-3}$ and are not too different
than the LIC measurement.  Likewise, the Gem Cloud is observed in the
second component of EX Hya ($n_e \sim 0.15$) and is similar to the LIC
measurement.  This is the first strong evidence that LISM clouds that
are dynamically distinct from the LIC nonetheless have a similar
electron density.  However, not all LISM clouds share similar electron
density properties.  The Hyades Cloud measurement observed in the
first component of G191-B2B is consistently measured at a high
electron density ($\sim$0.5) in all three techniques.  The Hyades
Cloud appears to be a decelerated cloud at the leading edge of the
platoon of LISM clouds beyond the LIC
\citep{redfield01,redfield07lism4}.  It is therefore closer to
G191-B2B than the LIC and without any obvious LISM clouds between it
and G191-B2B to shield the ionizing radiation.  The Hyades Cloud
appears to have an enhanced electron density due to the
photoionization of G191-B2B, while also shielding other LISM clouds,
such as the LIC, from the strong ionizing radiation of G191-B2B.  No
obvious measurements of the second largest LISM cloud and likely
future interstellar environment of the Sun, the G Cloud, exist in
this sample.

\subsection{Sources of Ionization and Self-Shielding}

Almost all of our background stars are white dwarfs or early-type
stars and therefore strong UV photon sources and significant
contributors to the local radiation field.  Indeed, \citet{vallerga98}
provided an inventory of 54 strong extreme-UV (EUV) stars that largely
determine the local radiation field.  The local EUV field is dominated
by the B star $\epsilon$ CMa and by three white dwarfs (Feige 24, HZ
43, and G191-B2B).  Both Feige 24 and G191-B2B are identified here as
having detectable \ion{C}{2}$^{\ast}$, while $\epsilon$ CMa is a
relatively distant star ($\sim$130 pc) and not included in our search
sample, and HZ 43 did not have any detectable \ion{C}{2}$^{\ast}$,
although this may be due to the significantly lower amount of LISM
material in that direction.  Three more of our 13 sight lines
($\alpha$ Aur, WD0050-332, and WD2309+105) are also included in the
radiation field inventory by \citet{vallerga98}, and we expect that
the other white dwarfs and early-type stars that make up our remaining
sight lines are also significant contributors to the local ionizing
radiation field.

In the left plot of Figure~\ref{fig:shield}, the electron density
measurements are shown as a function of the angle from $\epsilon$ CMa.
This is a test that was suggested in \citet{vallerga98}.
Unfortunately, most sight lines cluster around 90 degrees from the
direction of $\epsilon$ CMa, but nonetheless, no clear correlation is
detected.  This constrasts with a possible ionization gradient toward
$\epsilon$ CMa, based on hydrogen and helium column densities measured
in the extreme ultraviolet toward many nearby white dwarf stars
\citep{wolff99}.  Our measurements indicate that $\epsilon$ CMa does
not singularly dominate the ionization structure of the LISM, and
other contributors, such as the other nearby WDs and early type stars,
may be significant sources of the radiation field that dictates the
electron density in the LISM.

The right plot of Figure~\ref{fig:shield} demonstrates a possible test
of self-shielding in LISM clouds.  The figure plots the difference of
electron densities along the same sight line as a function of the
total column density along the line of sight.  Those with low columns
should show evidence of less shielding, while large columns will
provide significant shielding and a more dramatic difference in
electron density measurements.  No such correlation is seen,
indicating that for the entire LISM sample, there does not appear to
be pervasive shielding of ionizing radiation along each specific line
of sight.  This diagnostic assumes that for each sight line, the
dominant ionization source is the background star itself, which will
not necessarily be the case.  A more sophisticated three-dimensional
morphological model of the LISM that includes the effects of known
ionizing sources is needed to accurately determine if the observed
electron density is consistent with shielding of LISM clouds.
However, isolated sight lines of strong ionization sources may be used
to investigate the impact of shielding on electron density
measurements.

As mentioned in the previous section, the G191-B2B sight line provides
a possible example of self-shielding.  Because many of our background
targets are significant EUV sources, clouds closer to the source
(i.e., more distant from the Sun) should shield the clouds farther
from the source (i.e., nearer to the Sun).  The LIC components offer
excellent examples of this, since we know the LIC component is tracing
gas closest to the Sun.  The G191-B2B sight line, and others including
the $\alpha$ Gru, WD0232+035, and WD1210+533 sight lines, are cases
where the LIC component electron density is less than the other
observed component, presumably probing material from a more distant
cloud that lies closer to the ionizing source (assuming it is the
background star observed).  \citet{gry01} also noted this phenomena in
looking at the electron densities toward the most dominant ionization
source, $\epsilon$~CMa.  They identified two absorbers with local
clouds, the LIC and Blue Cloud, which had low electron densities, $n_e
= $0.08--0.17 and 0.016--0.088 cm$^{-3}$, respectively, while the
distant absorber, which was located closer to $\epsilon$~CMa, had a
significantly higher electron density, $n_e = $0.18--0.28 cm$^{-3}$.
Several counter-examples, such as GD659, and 74 Psc A and B, are also
observed, where the LIC does not have a significantly lower electron
density.  However, these are among the weakest ionizing sources in our
sample.  The 74 Psc stars are inactive and relatively cool A stars,
while GD659 is the coolest ($\sim$35000 K) of the singleton white
dwarfs in our sample, which are all $\gtrsim$50000 K
\citep{lajoie07,holberg06}.

\subsection{LISM Pressure Measurements}

The electron density is also important for its implication for
pressure measurements of LISM material.  This is a particularly
critical issue due to the apparent disparity of the warm LISM clouds,
such as those observed by \ion{C}{2}$^{\ast}$ absorption, which have
pressures, $P/k = nT \sim 3000$ K~cm$^{-3}$
\citep[e.g.,][]{redfield06,jenkins02}, and the hot, tenuous Local
Bubble gas that surrounds the warm clouds, which has pressures
$\sim$10000 K~cm$^{-3}$ \citep[e.g.,][]{snowden90}.  Recent evidence
of soft X-ray emission at the heliosphere due to charge exchange
between the solar wind and incoming LISM appears to contribute to the
soft X-ray emission that was previously fully attributed to the hot
Local Bubble gas \citep[e.g.,][]{lallement04}.  A revised inventory of
soft X-ray emission may lower the temperature and/or density of the
hot gas and reduce the pressure discrepancy.

The top axis of Figure~\ref{hist} shows the range of measured
pressures that we obtain from our electron density measurements.  This
calculation assumes temperature is constant for all sight lines, and a
simple photoionization relationship between the electron density
($n_e$) and the neutral hydrogen density ($n_{\rm HI}$).  In the case
of temperature, we know this assumption is not completely valid since
we see some variation about the mean LISM value \citep[$T = 6680$
K][]{redfield04sw,redfield04tt}.  However, since independent
temperature measurements are not available for the majority of sight
lines studied here, and since the dispersion about the mean
temperature is not high, it is a reasonable initial assumption.  We
assume $n_{\rm HI} = n_e^2 \alpha({\rm H})/\Gamma({\rm H})$ \citep[see
Equation 7 in][]{sofia98}.  The balance of the recombination rate
($\alpha$) and the ionization rate ($\Gamma$) are assumed to be
constant in the LISM and we calibrate this quantity such that the LISM
average electron density ($n_e$) is consistent with the LISM
measurement of $n_{\rm HI} = 0.222$ cm$^{-3}$, based on {\it in situ}
measurements of $n_{\rm HeI} = 0.0151$ cm$^{-3}$ \citep{gloeckler04},
and the \ion{H}{1} to \ion{He}{1} column density ratio observed toward
nearby white dwarfs \citep{dupuis95}.  We assume $n_p = n_{\rm e}$ and
$n_{\rm He} = 0.1 n_{\rm H}$.  This calculation breaks down at very
low densities ($n_e \leq 0.007$ cm$^{-3}$), where the derived hydrogen
density is less than the minimum value allowed from observations of
nearby stars, measured by dividing the observed hydrogen column
density by the distance to the background source
\citep{redfield04sw,linsky00}.  At high densities ($n_e \geq 0.4$
cm$^{-3}$), our assumption of a constant ionization rate fails
severely because the resulting column density through a typical LISM
cloud ($N_{\rm HI} \geq 10^{19.5}$) leads to significant shielding of
ionizing ratiation \citep{jenkins04}.  The top axis of
Figure~\ref{hist} is only printed for the range of densities for which
our calculation is reasonable.  The distribution of electron densities
then translates into an unweighted mean $P/k = 3300^{+5500}_{-1900}$
K~cm$^{-3}$, consistent with the range of values determined for other
nearby stars using excited transitions of \ion{C}{1} by
\citet{jenkins02}.

\section{Conclusions}\label{sions}

We analyze high spectral resolution observations of LISM absorption in
order to survey the electron density in nearby interstellar material.
These measurements should provide important constraints on the
ionization and abundance patterns of the LISM
\citep[e.g.,][]{slavin02,sofia98,jenkins00}, as well as on
evolutionary models of all phases of the LISM \citep{breitschwerdt06}.
A summary of our results is as follows:

\begin{enumerate}
\item We searched the entire {\it HST} spectroscopic database of
nearby stars ($<$100 pc) for detections of \ion{C}{2}$^{\ast}$.  Of
the $\sim$417 total nearby sight lines, we find 13 that show
\ion{C}{2}$^{\ast}$ absorption in 23 different velocity components.
The vast majority of these detections are new.

\item Using the \ion{C}{2}$^{\ast}$ to \ion{C}{2} ratio, we infer the
electron density.  To increase the accuracy of our results,
particularly in terms of measuring the column density of the saturated
\ion{C}{2} resonance line, we employ three analytical strategies: (a)
simultaneously fitting both the \ion{C}{2}$^{\ast}$ and \ion{C}{2}
profiles, allowing the optical thin \ion{C}{2}$^{\ast}$ line to
constrain the line width, (b) using independently derived temperatures
from comparison of line widths to constrain the acceptable range of
line widths for \ion{C}{2}, and (c) using easily measured \ion{S}{2}
column densities as a proxy for \ion{C}{2} column density.

\item The distribution of electron densities based on using \ion{S}{2}
as a proxy for \ion{C}{2} is similar to the distribution based on
carbon alone, while significantly tighter.  This is a promising
technique to avoid grossly overestimating the \ion{C}{2} column
density based on the saturated line profile.

\item We find the distribution of measured LISM electron densities
($n_e$) is consistent with a log-normal profile, with a mean
(unweighted) value of $n_e($\ion{C}{2}$_{\rm SII}) =
0.11^{+0.10}_{-0.05}$ cm$^{-3}$.

\item We assign individual velocity components to specific LISM clouds
based on kinematical and spatial properties.  In particular, the LIC
is probed by seven different sight lines, which all give roughly
identical electron density measurements.  The weighted mean value for
the LIC is $n_e = 0.12 \pm 0.04$ cm$^{-3}$.

\item Two clouds, the NGP and Gem clouds, show similar electron
density properties as the LIC.  The Hyades Cloud, a decelerated cloud
at the leading edge of the platoon of LISM clouds, has a significantly
higher electron density than the LIC.  Observed toward G191-B2B, the
high electron density may be caused by the lack of shielding from such
a strong radiation source.

\item Almost all of our background sources are significant ionizing
sources that may influence the ionization structure and thereby the
electron density of the gas along the line of sight.  We do not find
evidence that the ionization structure of the LISM is dominated by a
single source, namely $\epsilon$ CMa.

\item We see evidence of more distant clouds (i.e., those closest to
the ionizing sources) shielding nearer clouds (i.e., those farthest
from the ionizing sources).  In several examples, the LIC component
which is known to be farthest from the ionizing source has a lower
electron density than the component along the same line of sight that
is nearer to the radiation source.  Although counter-examples exist,
they are toward the weakest radiation sources in our sample.

\item The range in electron density is used to estimate the range of
pressures that may be found in warm LISM clouds.  Given simple
assumptions, the measured electron densities correspond to an
unweighted mean pressure $P/k = 3300^{+5500}_{-1900}$ K cm$^{-3}$.

\end{enumerate}

\acknowledgements 
The authors would like to thank Jeff Linsky and Brian Wood for reading
the draft and providing helpful comments.  We thank the referee for a
careful reading and many excellent suggestions.  S.R. would like to
acknowledge support provided by NASA through Hubble Fellowship grant
HST-HF-01190.01 awarded by the Space Telescope Science Institute,
which is operated by the Association of Universities for Research in
Astronomy, Inc., for NASA, under contract NAS 5-26555.  This research
has made use of NASA's Astrophysics Data System Bibliographic
Services.  This research has made use of the SIMBAD database, operated
at CDS, Strasbourg, France.  Some of the data presented in this paper
were obtained from the Multimission Archive at the Space Telescope
Science Institute (MAST). STScI is operated by the Association of
Universities for Research in Astronomy, Inc., under NASA contract
NAS5-26555. Support for MAST for non-HST data is provided by the NASA
Office of Space Science via grant NAG5-7584 and by other grants and
contracts.

{\it Facilities:} \facility{HST (GHRS, STIS)}


\begin{deluxetable}{ccccrrc}
\tablewidth{0pt}
\tablecaption{Stellar Parameters\label{starpar}}
\tablehead{
\colhead{HD} & \colhead{Other} & \colhead{Other} & \colhead{Spectral} &
\colhead{$l$} & \colhead{$b$} & \colhead{Distance\tablenotemark{a}}
\\
\colhead{\#} & \colhead{Name} & \colhead{Name} & \colhead{Type} &
\colhead{(deg)} & \colhead{(deg)} & \colhead{(pc)}
}
\startdata
34029 & $\alpha$ Aur & Capella & G0III+G8III &   162.58 &  $+$4.566 &
12.94 $\pm$
0.15
\\
120315 & $\eta$ UMa & Alcaid & B3V &   100.69 &  $+$65.32 & 30.87 $\pm$
0.71
\\
209952 & $\alpha$ Gru & Alnair & B7IV & 349.99 & $-$52.47 & 31.10 $\pm$
0.79
\\
\nodata & WD 0050--332 & GD 659 & DA1\tablenotemark{b} & 299.14 & $-$84.11
&
$\sim$58
\\
\nodata &  EX Hya & \nodata & M5/M6+DA & 303.18 & $+$33.62 & 64.5 $\pm$
1.2
\\
6457 & 74 Psc B & \nodata & A0V & 127.34 & $-$41.28 & 68.2 $\pm$ 8.0
\\
\nodata & WD 0501+527 & G191-B2B & DA1\tablenotemark{b} & 155.95 &
$+$7.099 & 69
$\pm$15
\\
6456 & 74 Psc A & \nodata & A1V & 127.34 & $-$41.27 & 73.2 $\pm$ 7.9
\\
\nodata & WD 0232+035 & Feige 24 & DAZQO1\tablenotemark{b} & 165.96 &
$-$50.26 & 74
$\pm$ 20
\\
\nodata & WD 2309+105 & GD 246 & DA1\tablenotemark{b} & 87.262 & $-$45.11
&
$\sim$79
\\
\nodata & WD 1210+533 & \nodata & DAO\tablenotemark{b} & 135.61 & $+$63.11
&
$\sim$87
\\
128345 & $\rho$ Lup & \nodata & B5V & 320.13 & $+$9.857 & 95.1 $\pm$ 6.4
\\
\nodata & IX Vel & \nodata & B8V+DA & 264.92 & $-$7.890 & 96.3 $\pm$ 9.1
\enddata
\tablecomments{All values from SIMBAD unless otherwise noted.}
\tablenotetext{a}{All distances are {\it Hipparcos} distances
\citep{perryman97} except for WD~0050-332 and WD~2309+105 whose distances
are from \citet{vennes97} and WD~1210+533 whose distance is from
\citet{holberg98}.}
\tablenotetext{b}{White dwarf spectral types taken from \citet{mccook99}.}
\end{deluxetable}

\begin{deluxetable}{cccccccc}
\tabletypesize{\tiny}
\tablewidth{0pt}
\tablecaption{Observational Parameters\label{obspar}}
\tablehead{
\colhead{} & \colhead{} & \colhead{} & \colhead{} & \colhead{Spectral} &
\colhead{} & \colhead{Exposure} & \colhead{}
\\
\colhead{Other} & \colhead{Other} & \colhead{} & \colhead{} &
\colhead{Range}
& \colhead{Resolution} & \colhead{Time\tablenotemark{a}} & \colhead{}
\\
\colhead{Name} & \colhead{Name} & \colhead{Instrument} &
\colhead{Grating} & \colhead{(\AA)} & \colhead{($\lambda/\Delta\lambda$)}
& \colhead{($s$)} & \colhead{Dataset}
}
\startdata
$\alpha$ Aur & Capella & GHRS & ECH-A & 1331$-$1338 & 100000 & 761.6 &
Z2UW030BT
\\
$\eta$ UMa & Alcaid & GHRS & ECH-A & 1332$-$1339 & 100000 & 1171.4 &
Z3CL020ET
\\
 & & GHRS & ECH-B & 2790$-$2805 & 100000 & 108.8 & Z3CL030HT
\\
 & & GHRS & ECH-B & 2845$-$2859 & 100000 & 652.8 & Z3CL0308T
\\
 & & & & & & & Z3CL0307T
\\
$\alpha$ Gru & Alnair & GHRS & G160M & 1309$-$1345 & 20000 & 230.4 &
Z1720109T
\\
 & & GHRS & ECH-B & 2791$-$2806 & 100000 & 108.8 & Z1720209T
\\
 & & GHRS & ECH-B & 2847$-$2861 & 100000 & 217.6 & Z172020AT
\\
WD 0050-332 & GD 659 & STIS & E140H & 1170$-$1372 & 114000 & 4134 &
O4G101010
\\
 & & & & & & & O4G101020
\\
EX Hya & \nodata & STIS & E140M & 1140$-$1735 & 45800 & 15200 & O68301010
\\
 & & & & & & & O68301020
\\
 & & & & & & & O68301030
\\
 & & & & & & & O68302010
\\
 & & & & & & & O68302020
\\
 & & & & & & & O68302030
\\
74 Psc B & HR 311 & STIS & E140H & 1170$-$1372 & 114000 & 2768 & O56L02010
\\
 & & & & & & & O56L02020
\\
 & & & & & & & O56L02030
\\
WD 0501+527 & G191-B2B & STIS & E140H & 1170$-$1517\tablenotemark{b} &
114000 & 5623 & O57U01020
\\
 & & & & & & & O6HB10040
\\
 & & & & & & & O6HB10050
\\
 & & & & & & & O6HB10060
\\
 & & & & & & & O6HB10070
\\
 & & & & & & & O6HB10080
\\
 & & & & & & & O6HB10090
\\
 & & STIS & E230H & 2624$-$3095\tablenotemark{b} & 114000 & 6013 &
O6HB30080
\\
 & & & & & & & O6HB30090
\\
 & & & & & & & O6HB300B0
\\
 & & & & & & & O6HB300C0
\\
 & & & & & & & O6HB300D0
\\
 & & & & & & & O6HB300E0
\\
74 Psc A & HR 310 & STIS & E140H & 1170$-$1372 & 114000 & 4128 & O56L01010
\\
 & & & & & & & O56L01020
\\
 & & & & & & & O56L01030
\\
 & & & & & & & O56L51020
\\
WD 0232+035 & Feige 24 & STIS & E140M & 1150$-$1735 & 45800 & 4176 &
O4G701010
\\
 & & & & & & & O4G702010
\\
WD 2309+105 & GD 246 & STIS & E140H & 1170$-$1372 & 114000 & 2420 &
O4G102020
\\
WD 1210+533 & \nodata & STIS & E140M & 1140$-$1735 & 45800 & 8371 &
O5F203010
\\
 & & & & & & & O5F203020
\\
 & & & & & & & O5F204010
\\
$\rho$ Lup & HR 5453 & STIS & E140M & 1150$-$1735 & 45800 & 1900 &
O8S602010
\\
IX Vel & \nodata & STIS & E140M & 1140$-$1735 & 45800 & 5250 & O5BI01010
\\
 & & & & & & & O5BI02010
\\
 & & & & & & & O5BI03010
\enddata
\tablenotetext{a}{For targets with multiple datasets, the exposure time
listed is the sum of the exposure times for each dataset.}
\tablenotetext{b}{Spectral range encompassed by multiple datasets.}
\end{deluxetable}

\begin{deluxetable}{cccrcccc}
\tablewidth{0pt}
\tabletypesize{\scriptsize}
\tablecaption{Fit Parameters for \ion{C}{2} ISM Velocity
Components\label{fitpar}}
\tablehead{
\colhead{Other} & \colhead{Other} & \colhead{Comp.} & \colhead{$v$} &
\colhead{$b$} & \colhead{log$N$(C\,{\tiny II})\tablenotemark{a}} &
\colhead{log$N$(C\,{\tiny II}$_{{\rm SII}}$)} &
\colhead{log$N$(C\,{\tiny II}$^{\ast}$)\tablenotemark{b}}
\\
\colhead{Name} & \colhead{Name} & \colhead{$\#$} & \colhead{(km s$^{-1}$)}
&
\colhead{(km s$^{-1}$)} & \colhead{log (cm$^{-2}$)} &
\colhead{log (cm$^{-2}$)} & \colhead{log (cm$^{-2}$)}
}
\startdata
$\alpha$ Aur & Capella & 1 & 20.78 $\pm$ 0.28 &
3.48$^{+0.15}_{-0.19}$\tablenotemark{c} & 14.67$_{-0.22}^{+0.14}$ &
\nodata & 12.62$_{-0.07}^{+0.07}$ \vspace{.5mm}
\\
$\eta$ UMa & Alcaid & 1 & $-$2.09 $\pm$ 0.24 &
3.76$^{+0.09}_{-0.11}$\tablenotemark{c} & 14.43$_{-0.14}^{+0.12}$ &
\nodata & 12.39$_{-0.02}^{+0.02}$ \vspace{.5mm}
\\
 & & 2 & 9.72 $\pm$ 0.95 & 5.60$^{+0.09}_{-0.11}$\tablenotemark{c} &
13.12$_{-0.07}^{+0.06}$ & \nodata & 11.60$_{-0.05}^{+0.05}$ \vspace{.5mm}
\\
$\alpha$ Gru & Alnair & 1 & $-$22.50 $\pm$ 0.60\tablenotemark{d} & 3.35
$\pm$ 0.73 & 14.98$_{-0.19}^{+0.19}$ & \nodata & 12.69$_{-0.17}^{+0.14}$
\vspace{.5mm}
\\
 & & 2 & $-$12.63 $\pm$ 0.60\tablenotemark{d} & 3.49 $\pm$ 0.51 &
14.52$_{-0.23}^{+0.18}$ & \nodata & 12.73$_{-0.17}^{+0.17}$ \vspace{.5mm}
\\
 & & 3 & $-$8.32 $\pm$ 0.60\tablenotemark{d} & 3.9 $\pm$ 2.2 &
13.63$_{-0.23}^{+0.23}$ & \nodata & 12.17$_{-0.48}^{+0.48}$ \vspace{.5mm}
\\
WD 0050-332 & GD 659 & 1 & 6.17 $\pm$ 0.38 & 4.31 $\pm$ 0.77 &
15.12$_{-0.32}^{+0.32}$ & 14.808$_{-0.065\;-0.216}^{+0.065\;+0.216}$ &
12.71$_{-0.05}^{+0.05}$ \vspace{.5mm}
\\
 & & 2 & 12.3 $\pm$ 1.0 & 4.67 $\pm$ 0.67 & 15.12$_{-0.29}^{+0.29}$ &
14.062$_{-0.350\;-0.406}^{+0.240\;+0.317}$ & $<$11.9
\\
EX Hya & \nodata & 1 & $-$15.2 $\pm$ 1.2 & 5.03 $\pm$ 0.72 &
16.64$_{-0.27}^{+0.27}$ & 15.232$_{-0.120\;-0.239}^{+0.120\;+0.239}$ &
13.01$_{-0.10}^{+0.10}$ \vspace{.5mm}
\\
 & & 2 & $-$6.2 $\pm$ 2.3 & 6.1 $\pm$ 1.9 & 14.75$_{-0.53}^{+0.53}$ &
14.792$_{-0.290\;-0.356}^{+0.290\;+0.356}$ & 12.78$_{-0.17}^{+0.17}$
\vspace{.5mm}
\\
74 Psc B & HR 311 & 1 & $-$5.75 $\pm$ 0.50 & 4.00 $\pm$ 0.93 &
16.03$_{-0.16}^{+0.16}$ & 15.444$_{-0.017\;-0.207}^{+0.017\;+0.207}$ &
13.07$_{-0.02}^{+0.02}$ \vspace{.5mm}
\\
 & & 2 & 10.47 $\pm$ 0.70 & 4.26 $\pm$ 0.44 & 15.46$_{-0.25}^{+0.25}$ &
14.744$_{-0.052\;-0.213}^{+0.052\;+0.213}$ & 12.82$_{-0.03}^{+0.03}$
\vspace{.5mm}
\\
WD 0501+527 & G191-B2B & 1 & 5.98 $\pm$ 0.17 &
4.10$^{+0.29}_{-0.31}$\tablenotemark{c} & 14.8$_{-1.2}^{+0.3}$ &
14.345$_{-0.087\;-0.224}^{+0.087\;+0.224}$ & 13.14$_{-0.02}^{+0.02}$
\vspace{.5mm}
\\
 & & 2 & 16.98 $\pm$ 0.60 & 3.43$^{+0.21}_{-0.26}$\tablenotemark{c} &
15.42$_{-0.17}^{+0.17}$ & 14.550$_{-0.052\;-0.213}^{+0.052\;+0.213}$ &
12.28$_{-0.09}^{+0.09}$ \vspace{.5mm}
\\
74 Psc A & HR 310 & 1 & $-$5.97 $\pm$ 0.52 & 3.82 $\pm$ 0.59 &
16.42$_{-0.14}^{+0.14}$ & 15.399$_{-0.041\;-0.211}^{+0.038\;+0.210}$ &
13.07$_{-0.03}^{+0.03}$ \vspace{.5mm}
\\
 & & 2 & 10.29 $\pm$ 0.76 & 4.75 $\pm$ 0.73 & 15.04$_{-0.25}^{+0.25}$ &
14.762$_{-0.110\;-0.234}^{+0.110\;+0.234}$ & 12.96$_{-0.04}^{+0.04}$
\vspace{.5mm}
\\
WD 0232+035 & Feige 24 & 1 & 3.81 $\pm$ 0.47 & 4.36 $\pm$ 0.53 &
15.18$_{-0.42}^{+0.42}$ & 14.877$_{-0.087\;-0.224}^{+0.072\;+0.219}$ &
13.00$_{-0.02}^{+0.02}$ \vspace{.5mm}
\\
 & & 2 & 17.4 $\pm$ 4.6 & 3.1 $\pm$ 1.4 & 15.96$_{-0.55}^{+0.45}$ &
13.812$_{-0.380\;-0.432}^{+0.380\;+0.432}$ & $<$11.8
\\
WD 2309+105 & GD 246 & 1 & $-$9.7 $\pm$ 2.5\tablenotemark{e} & 4.20 $\pm$
0.34 & 15.16$_{-0.30}^{+0.30}$ &
15.321$_{-0.035\;-0.209}^{+0.035\;+0.209}$ & 13.05$_{-0.04}^{+0.03}$
\vspace{.5mm}
\\
 & & 2 & 0.3 $\pm$ 2.5\tablenotemark{e} & 5.10 $\pm$ 0.74 &
14.18$_{-0.26}^{+0.53}$ & 14.716$_{-0.099\;-0.229}^{+0.090\;+0.225}$ &
$<$12.3
\\
WD 1210+533 & \nodata & 1 & $-$23.4 $\pm$ 1.2 & 3.6 $\pm$ 1.1 &
13.40$_{-0.13}^{+0.13}$ & $<$14.3 & $<$12.0
\\
 & & 2 & $-$9.0 $\pm$ 1.3 & 4.3 $\pm$ 1.1 & 15.50$_{-0.46}^{+0.46}$ &
14.632$_{-0.250\;-0.324}^{+0.340\;+0.398}$ & 13.00$_{-0.16}^{+0.16}$
\vspace{.5mm}
\\
 & & 3 & $-$2.3 $\pm$ 2.0 & 4.8 $\pm$ 1.2 & 15.44$_{-0.69}^{+0.69}$ &
15.272$_{-0.120\;-0.239}^{+0.120\;+0.239}$ & 13.06$_{-0.30}^{+0.30}$
\vspace{.5mm}
\\
$\rho$ Lup & HR 5453 & 1 & $-$16.11\tablenotemark{f} & 5.17 $\pm$ 0.74 &
15.80$_{-0.43}^{+0.43}$ & 15.642$_{-0.100\;-0.229}^{+0.100\;+0.229}$ &
13.27$_{-0.17}^{+0.17}$ \vspace{.5mm}
\\
 & & 2 & $-$9.1\tablenotemark{f} & 4.3 $\pm$ 1.1 & 15.14$_{-0.61}^{+0.61}$
& 15.012$_{-0.310\;-0.372}^{+0.310\;+0.372}$ & 13.08$_{-0.31}^{+0.31}$
\vspace{.5mm}
\\
IX Vel & \nodata & 1 & 4.9 $\pm$ 1.2 & 5.73 $\pm$ 0.61 &
14.39$_{-0.24}^{+0.24}$ & 14.700$_{-0.140\;-0.249}^{+0.140\;+0.249}$ &
12.64$_{-0.06}^{+0.06}$ \vspace{.5mm}
\\
 & & 2 & 19.08 $\pm$ 0.24 & 4.17 $\pm$ 0.38 & 16.39$_{-0.14}^{+0.14}$ &
15.402$_{-0.023\;-0.208}^{+0.024\;+0.208}$ &
13.43$_{-0.03}^{+0.03}$ \vspace{.5mm}
\enddata
\tablenotetext{a}{ Resonance line parameters derived from simultaneous
fits.}
\tablenotetext{b}{ Excited line parameters are weighted means of
simultaneous and excited-only fits.}
\tablenotetext{c}{ Fixed $b$ values based on independent temperature and
turbulent velocity measurements \citep{redfield04tt}.}
\tablenotetext{d}{ Fixed velocity difference between components based
on measurements of interstellar Fe\,{\tiny II} and Mg\,{\tiny II}
\citep{redfield02}.}
\tablenotetext{e}{ Fixed velocity difference between components based on
measurements of interstellar S\,{\tiny II}.}
\tablenotetext{f}{ Fixed velocities based on measurements of interstellar
S\,{\tiny II}.}
\end{deluxetable}

\begin{deluxetable}{cccrcc}
\tablewidth{0pt}
\tabletypesize{\scriptsize}
\tablecaption{Fit Parameters for \ion{S}{2} ISM Velocity Components\label{tab:s2fitpar}}
\tablehead{
\multicolumn{3}{c}{} & \multicolumn{3}{c}{Resonance Line\tablenotemark{a}} 
\\
\cline{4-6} \colhead{Other} & \colhead{Other} & \colhead{Comp.} & \colhead{$v$} & \colhead{$b$} & \colhead{log$N$} 
\\
\colhead{Name} & \colhead{Name} & \colhead{$\#$} & \colhead{(km s$^{-1}$)} & \colhead{(km s$^{-1}$)} & \colhead{log (cm$^{-2}$)} 
}
\startdata
WD 0050-332 & GD 659 & 1 & 7.00 $\pm$ 0.90 & 3.07 $\pm$ 0.67 & 13.726 $\pm$ 0.065
\\
 & & 2 & 15.0 $\pm$ 2.9 & 2.8 $\pm$ 2.5 & 12.98$_{-0.35}^{+0.24}$ 
\\
EX Hya & \nodata & 1 & $-$12.7 $\pm$ 2.1 & 7.4 $\pm$ 2.9 & 14.15 $\pm$ 0.12
\\
                & & 2 & $-$4.1 $\pm$ 4.0  & 2.6 $\pm$ 1.9 & 13.71 $\pm$ 0.29
\\
74 Psc B & HR 311 & 1 & $-$5.59 $\pm$ 0.93 & 3.48 $\pm$ 0.18 & 14.362 $\pm$ 0.017
\\
              & & 2 & 11.8 $\pm$ 1.4 & 3.91 $\pm$ 0.65 & 13.662 $\pm$ 0.052
\\
WD 0501+527 & G191-B2B & 1 & 8.4  $\pm$ 1.2 & 2.21 $\pm$ 0.71 & 13.263 $\pm$ 0.087
\\
                 & & 2 & 20.0 $\pm$ 1.6 & 4.46 $\pm$ 0.92 & 13.468 $\pm$ 0.052
\\
74 Psc A & HR 310 & 1 & $-$6.21 $\pm$ 0.92 & 4.19 $\pm$ 0.31 & 14.317$_{-0.041}^{+0.038}$ 
\\
              & & 2 & 12.3 $\pm$ 2.5 & 4.6 $\pm$ 1.2 & 13.68 $\pm$ 0.11
\\
WD 0232+035 & Feige 24 & 1 & 4.11 $\pm$ 0.64 & 4.18 $\pm$ 0.95 & 13.795$_{-0.087}^{+0.072}$ 
\\
                    & & 2 & 16.8 $\pm$ 3.8 & 2.6 $\pm$ 2.3 & 12.73 $\pm$ 0.38
\\
WD 2309+105 & GD 246 & 1 & $-$8.3 $\pm$ 1.0 & 2.42 $\pm$ 0.21 & 14.239 $\pm$ 0.035 
\\
 & & 2 & 1.6 $\pm$ 1.5 & 4.9 $\pm$ 2.2 & 13.634$_{-0.099}^{+0.090}$ 
\\
WD 1210+533 & \nodata & 1 & \nodata           &  \nodata      & $<$13.2
\\
        &             & 2 & $-$12.0 $\pm$ 3.6 & 5.0 $\pm$ 2.6 & 13.55$_{-0.25}^{+0.34}$
\\
 & &                    3 & $-$3.10 $\pm$ 0.94 & 3.0 $\pm$ 1.0 & 14.19 $\pm$ 0.12
\\
$\rho$ Lup & HR 5453 & 1 & $-$16.11 $\pm$ 0.79 & 5.22 $\pm$ 0.66 & 14.56 $\pm$ 0.10
\\
                  & & 2 & $-$9.1 $\pm$ 2.4 & 4.5 $\pm$ 2.0 & 13.93 $\pm$ 0.31
\\
IX Vel & \nodata & 1 & 5.3 $\pm$ 1.2 & 3.8 $\pm$ 1.2 & 13.618 $\pm$ 0.14
\\
               & & 2 & 20.37 $\pm$ 0.53 & 4.64 $\pm$ 0.41 & 14.320$_{-0.023}^{+0.024}$ 
\enddata
\tablenotetext{a}{ Resonance line parameters derived from simultaneous fits.}
\end{deluxetable}

\begin{deluxetable}{cccrcccrcc}
\tablewidth{0pt}
\tabletypesize{\tiny}
\tablecaption{Fit Parameters for \ion{Mg}{2} and \ion{Mg}{1} ISM Velocity Components\label{tab:mgfitpar}}
\tablehead{
\multicolumn{3}{c}{} & \multicolumn{3}{c}{Mg II Line\tablenotemark{a}} & \colhead{} & \multicolumn{3}{c}{Mg I Line}
\\
\cline{4-6} \cline{8-10} \colhead{Other} & \colhead{Other} & \colhead{Comp.} & \colhead{$v$} & \colhead{$b$} & \colhead{log$N$} & \colhead{} & \colhead{$v$} & \colhead{$b$} & \colhead{log$N$}
\\
\colhead{Name} & \colhead{Name} & \colhead{$\#$} & \colhead{(km s$^{-1}$)} & \colhead{(km s$^{-1}$)} & \colhead{log (cm$^{-2}$)} & \colhead{} & \colhead{(km s$^{-1}$)} & \colhead{(km s$^{-1}$)} & \colhead{log (cm$^{-2}$)}
}
\startdata
$\eta$ UMa & Alcaid & 1 & $-$2.271 $\pm$ 0.077 & 2.59 $\pm$ 0.18 & 12.653 $\pm$ 0.014 &               & $-$2.36 $\pm$ 0.30   & 2.59 $\pm$ 0.42 & 10.153 $\pm$ 0.04 \vspace{.5mm}
\\
 & & 2  & 3.5 $\pm$ 1.2 & 8.4 $\pm$ 3.5 & 11.27$^{+0.11}_{-0.14}$ &                   & 3.5 $\pm$ 1.8 & 4.8 $\pm$ 1.8 &  9.27$^{+0.13}_{-0.18}$ \vspace{.5mm}
\\
$\alpha$ Gru & Alnair & 1 & $-$24.05 $\pm$ 0.37 & 3.54 $\pm$ 0.31 & 11.659$^{+0.048}_{-0.053}$        & &   \nodata           & \nodata         & $<$9.1 \vspace{.5mm}
\\
 & & 2 & $-$14.98 $\pm$ 0.44 & 3.37 $\pm$ 0.24 & 13.463$^{+0.17}_{-0.27}$ &          & $-$15.09 $\pm$ 0.75 & 3.43 $\pm$ 0.77 & 10.99$^{+0.15}_{-0.24}$ \vspace{.5mm}
\\
 & & 3 & $-$7.59 $\pm$ 0.11 & 1.96 $\pm$ 0.37 & 12.83$^{+0.26}_{-0.31}$ &            & $-$7.6 $\pm$ 3.4   & 1.9$^{+2.3}_{-1.9}$ & 10.16$^{+0.27}_{-0.90}$ \vspace{.5mm}
\\
WD 0501+527 & G191-B2B & 1 & 8.22 $\pm$ 0.23  & 3.32 $\pm$ 0.34 & 13.56$_{-0.12}^{+0.29}$              & & 7.94 $\pm$ 0.56  & 3.38 $\pm$ 0.40 & 11.193 $\pm$ 0.029\vspace{.5mm}
\\
                 & & 2 & 18.81 $\pm$ 0.24 & 2.65 $\pm$ 0.11 & 12.714$_{-0.052}^{+0.046}$ &           & \nodata          & \nodata         & $<$10.1 \vspace{.5mm}
\\
\enddata
\tablenotetext{a}{ Resonance line parameters derived from simultaneous fits.}
\end{deluxetable}

\begin{deluxetable}{cccr@{}lr@{}l@{}lr@{}ll}
\tabletypesize{\scriptsize}
\tablewidth{0pt}
\tablecaption{Electron Densities\label{electron}}
\tablehead{
\colhead{Other} & \colhead{Other} & \colhead{Comp.} & \multicolumn{2}{c}{$n_{e}$} &
\multicolumn{3}{c}{$n_e$(\ion{C}{2}$_{{\rm SII}}$)\tablenotemark{a}} & \multicolumn{2}{c}{$n_e($\ion{Mg}{2}$/$\ion{Mg}{1}$)$\tablenotemark{b}} & \colhead{Cloud\tablenotemark{c}}
\\
\colhead{Name} & \colhead{Name} & \colhead{$\#$} & \multicolumn{2}{c}{(cm$^{-3}$)} &
\multicolumn{3}{c}{(cm$^{-3}$)} & \multicolumn{2}{c}{(cm$^{-3}$)} & \colhead{}
}
\startdata
$\alpha$ Aur & Capella & 1 & 0.140 & $_{-0.059}^{+0.060}$\tablenotemark{d} & \multicolumn{3}{c}{\nodata} & \multicolumn{2}{c}{\nodata} & LIC
\\
$\eta$ UMa & Alcaid & 1 & 0.165 & $_{-0.047}^{+0.051}$ & \multicolumn{3}{c}{\nodata} & 0.089 & $_{-0.089}^{+0.118}$ & NGP
\\
 & & 2 & 0.154 & $_{-0.027}^{+0.029}$ & \multicolumn{3}{c}{\nodata} & 0.085 & $_{-0.085}^{+0.217}$ &\nodata
\\
$\alpha$ Gru & Alnair & 1 & 0.081 & $_{-0.038}^{+0.053}$ & \multicolumn{3}{c}{\nodata} & $<$0.21 & & \nodata
\\
 & & 2 & 0.25 & $_{-0.13}^{+0.18}$ & \multicolumn{3}{c}{\nodata} & 0.28 & $_{-0.28}^{+0.51}$ & (Mic, Vel)
\\
 & & 3 & 0.5 & $_{-0.4}^{+1.1}$ & \multicolumn{3}{c}{\nodata} & 0.16 & $_{-0.16}^{+0.35}$\tablenotemark{d} & LIC
\\
WD 0050-332 & GD 659 & 1 & 0.060 & $_{-0.033}^{+0.068}$ &
0.12 & $_{-0.02}^{+0.03}$ & $_{-0.05}^{+0.08}$\tablenotemark{d} & \multicolumn{2}{c}{\nodata} & LIC, (Cet)
\\
 & & 2 & $<$0.0094 & & $<$0.11 & & & \multicolumn{2}{c}{\nodata} & (Vel)
\\
EX Hya & \nodata & 1 & 0.004 & $_{-0.005}^{+0.003}$ &
0.094 & $_{-0.03}^{+0.04}$ & $_{-0.044}^{+0.074}$ & \multicolumn{2}{c}{\nodata} & NGP, (Leo, G)
\\
 & & 2 & 0.17 & $_{-0.13}^{+0.40}$ & 0.15 & $_{-0.09}^{+0.16}$ & $_{-0.10}^{+0.21}$ & \multicolumn{2}{c}{\nodata} &
Gem, (Leo, Aur)
\\
74 Psc B & HR 311 & 1 & 0.017 & $_{-0.005}^{+0.008}$ &
0.067 & $_{-0.004}^{+0.004}$ & $_{-0.026}^{+0.041}$ & \multicolumn{2}{c}{\nodata} & \nodata
\\
 & & 2 & 0.036 & $_{-0.046}^{+0.028}$ & 0.19 & $_{-0.02}^{+0.03}$ & $_{-0.07}^{+0.12}$\tablenotemark{d}
& \multicolumn{2}{c}{\nodata} & LIC, (Hyades, Eri)
\\
WD 0501+527 & G191-B2B & 1 & 0.30 & $_{-0.28}^{+0.28}$ &
0.80 & $_{-0.15}^{+0.18}$ & $_{-0.33}^{+0.54}$ & 0.48 & $_{-0.48}^{+0.70}$ & Hyades
\\
 & & 2 & 0.011 & $_{-0.004}^{+0.006}$ &
0.081 & $_{-0.017}^{+0.021}$ & $_{-0.035}^{+0.055}$\tablenotemark{d} & $<$0.39 & & LIC
\\
74 Psc A & HR 310 & 1 & 0.007 & $_{-0.002}^{+0.003}$ &
0.073 & $_{-0.008}^{+0.009}$ & $_{-0.028}^{+0.046}$ & \multicolumn{2}{c}{\nodata} & \nodata
\\
 & & 2 & 0.13 & $_{-0.06}^{+0.10}$ & 0.25 & $_{-0.06}^{+0.07}$ & $_{-0.11}^{+0.18}$\tablenotemark{d} & \multicolumn{2}{c}{\nodata} &
LIC, (Hyades, Eri)
\\
WD 0232+035 & Feige 24 & 1 & 0.10 & $_{-0.06}^{+0.17}$ &
0.21 & $_{-0.04}^{+0.04}$ & $_{-0.08}^{+0.14}$ & \multicolumn{2}{c}{\nodata} & \nodata
\\
 & & 2 & $<$0.0011 & & $<$0.15 & & & \multicolumn{2}{c}{\nodata} & LIC, (G, Blue, Hyades)
\\
WD 2309+105 & GD 246 & 1 & 0.12 & $_{-0.06}^{+0.12}$ &
0.084 & $_{-0.009}^{+0.010}$ & $_{-0.033}^{+0.052}$ & \multicolumn{2}{c}{\nodata} & \nodata
\\
 & & 2 & $<$0.21 & & $<$0.060 & & & \multicolumn{2}{c}{\nodata} & (LIC, Eri)
\\
WD 1210+533 & \nodata & 1 & $<$0.62 & & $<$0.079 & & & \multicolumn{2}{c}{\nodata} & \nodata
\\
 & & 2 & 0.050 & $_{-0.036}^{+0.097}$ & 0.37 & $_{-0.20}^{+0.47}$ & $_{-0.22}^{+0.57}$
& \multicolumn{2}{c}{\nodata} & \nodata
\\
 & & 3 & 0.07 & $_{-0.06}^{+0.26}$ & 0.10 & $_{-0.05}^{+0.10}$ & $_{-0.06}^{+0.12}$\tablenotemark{d} & \multicolumn{2}{c}{\nodata} &
LIC
\\
$\rho$ Lup & HR 4353 & 1 & 0.046 & $_{-0.033}^{+0.082}$ &
0.067 & $_{-0.026}^{+0.037}$ & $_{-0.035}^{+0.057}$ & \multicolumn{2}{c}{\nodata} & (Gem)
\\
 & & 2 & 0.14 & $_{-0.12}^{+0.44}$ & 0.18 & $_{-0.13}^{+0.27}$ & $_{-0.14}^{+0.31}$ & \multicolumn{2}{c}{\nodata} &
(Gem)
\\
IX Vel & \nodata & 1 & 0.28 & $_{-0.12}^{+0.21}$ &
0.14 & $_{-0.04}^{+0.06}$ & $_{-0.06}^{+0.11}$ & \multicolumn{2}{c}{\nodata} & (G, Blue)
\\*
 & & 2 & 0.018 & $_{-0.005}^{+0.007}$ & 0.17 & $_{-0.01}^{+0.01}$ & $_{-0.07}^{+0.10}$
& \multicolumn{2}{c}{\nodata} & (Vel)
\enddata
\tablenotetext{a}{ Two errors listed: the first are based on the propagation of the column density errors only, while the second include errors in the cosmic abundances and the natural range of depletions of carbon and sulfur in the ISM.}
\tablenotetext{b}{ Assume LISM temperature appropriate for the line of sight based on multi-ion line widths or the LISM average \citep{redfield04tt}.  For the second component toward $\eta$~UMa, which \citet{redfield04tt} estimate a temperature of 0$^{+4400}_{-0}$~K, we use $T = 100$~K.}
\tablenotetext{c}{ In agreement with projected velocity and spatial
distribution \citep{redfield07lism4}.}
\tablenotetext{d}{ Used to calculate the weighted mean value for the LIC, $n_e({\rm LIC}) = 0.12 \pm 0.04$ cm$^{-3}$.}
\end{deluxetable}

\setcounter{figure}{2}

\begin{figure}
\epsscale{.7}
\plotone{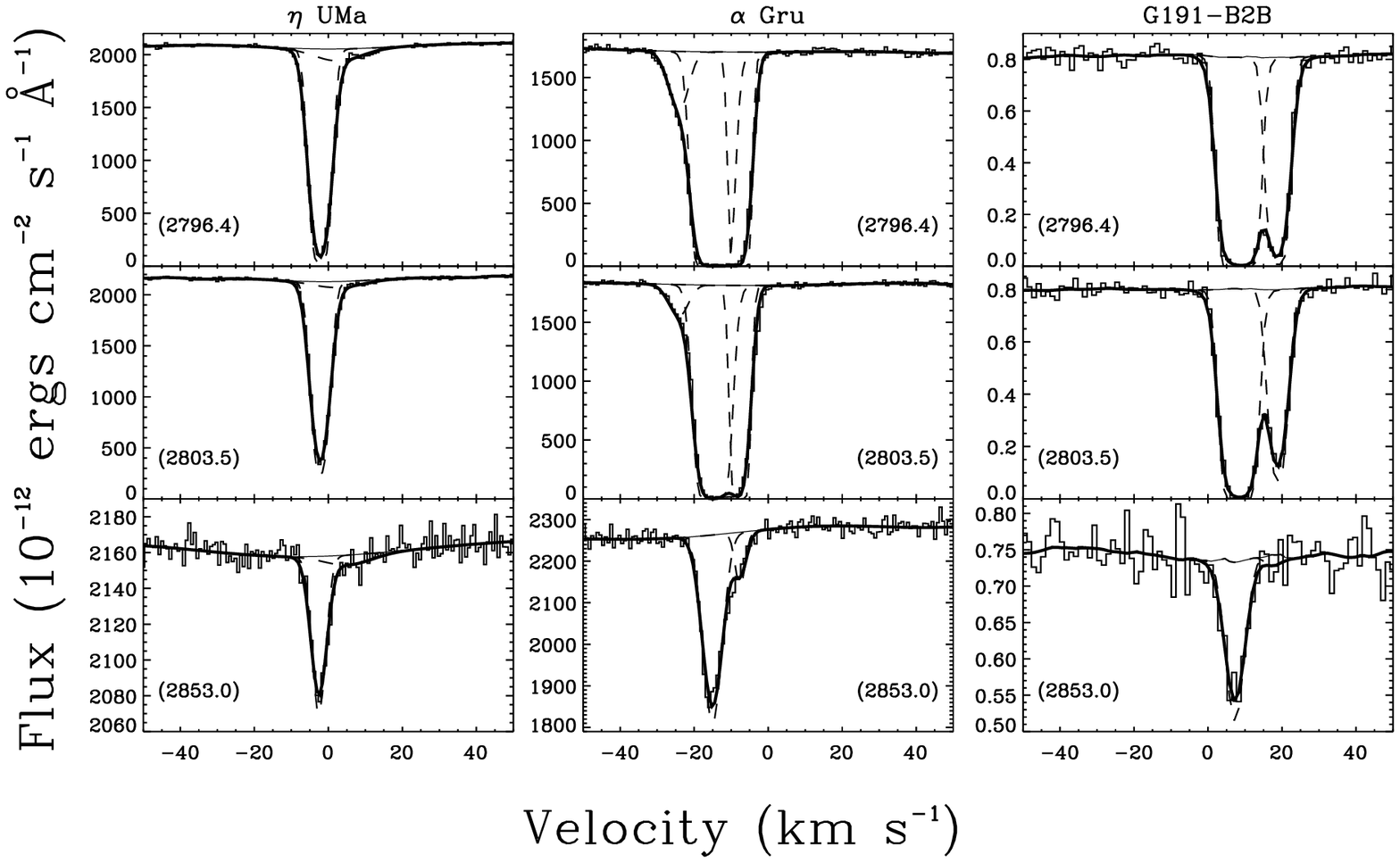}
\caption{Similar to Figure~\ref{9plot}, but for fits to interstellar
\ion{Mg}{1} and \ion{Mg}{2} absorption toward 3 of the 13 targets
which show \ion{C}{2}$^{\ast}$ absorption.  The ratio of ionization
stages are used to estimate the electron density for each
absorption component.  Both \ion{Mg}{2} lines are fit simultaneously.
The parameters for these fits are listed in
Table~\ref{tab:mgfitpar}. \label{fig:mgplot}}
\end{figure}

\begin{figure}
\epsscale{1.25}
\plotone{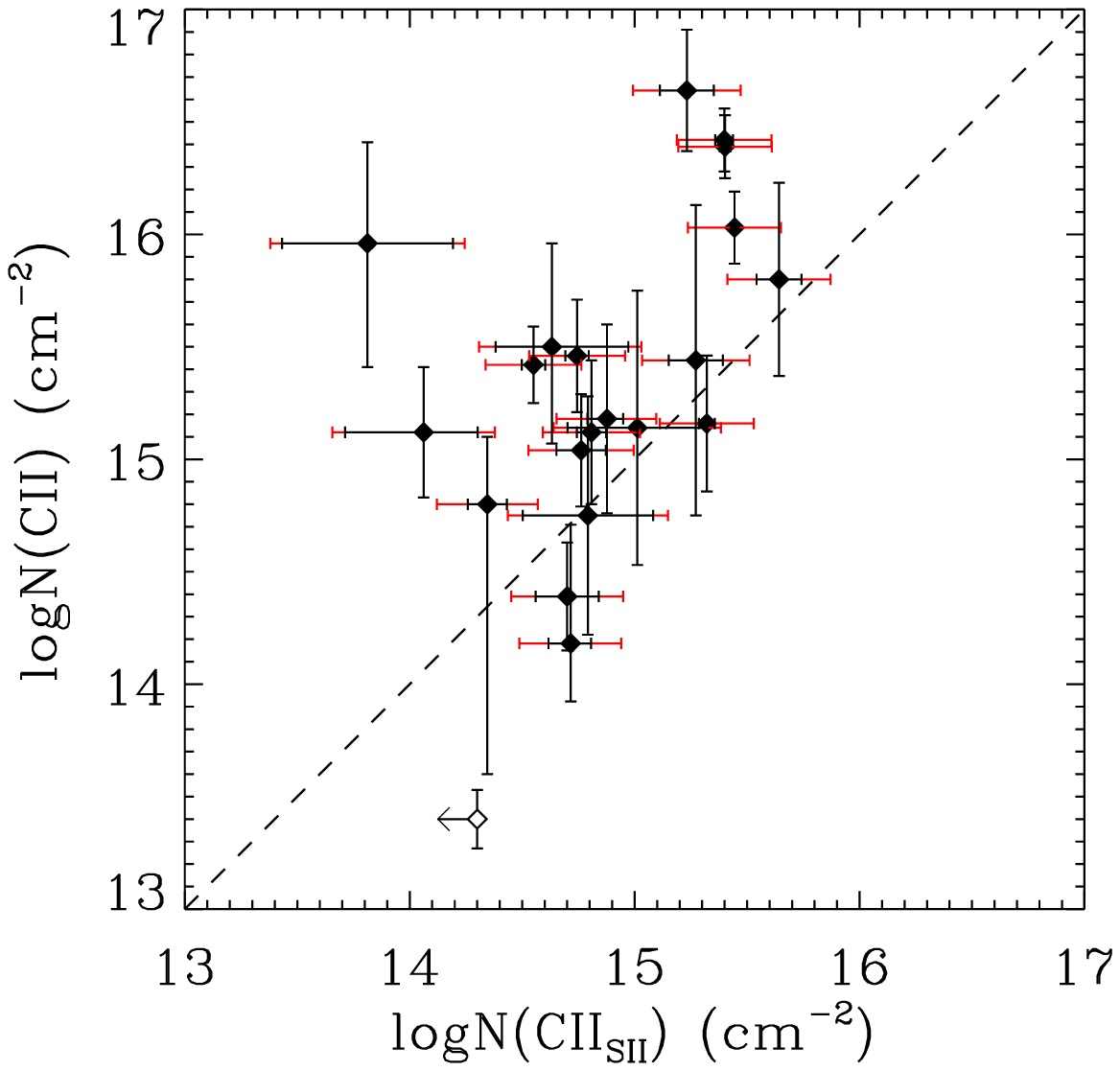}
\caption{Estimated \ion{C}{2} column density using \ion{S}{2} as a
proxy [$N$(\ion{C}{2}$_{\rm SII}$)] versus \ion{C}{2} resonance line
column density derived from the saturated lines directly
[$N$(\ion{C}{2})].  The systematic errors due to the conversion from
$N$(\ion{S}{2}) (solid red lines) extend beyond the random \ion{S}{2}
fitting errors.  Unity (dashed line) bisects the plot
window. \label{fig:cscomp}}
\end{figure}

\begin{figure}
\epsscale{.8}
\plotone{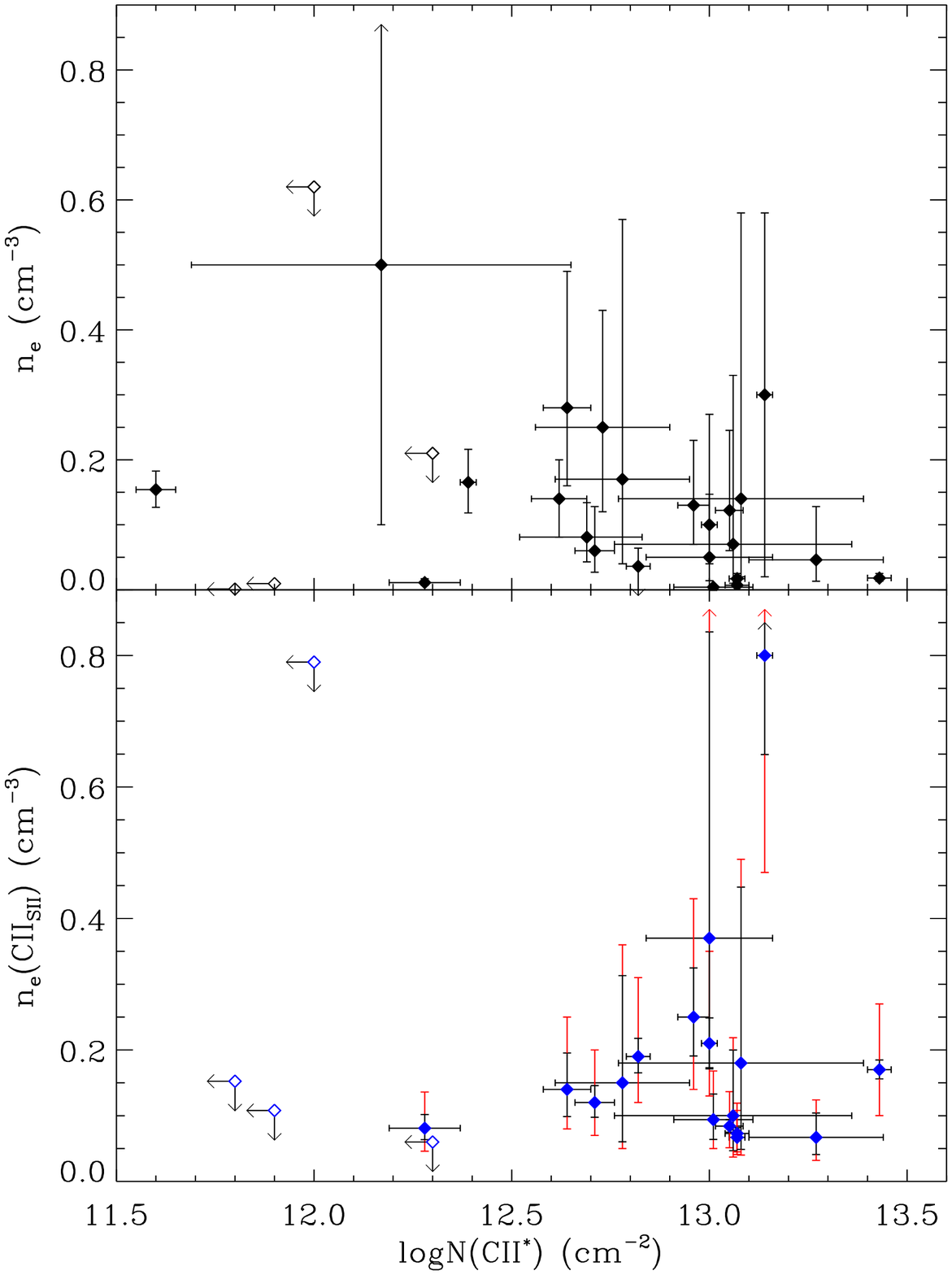}
\caption{Excited line column density versus electron density
(top, black) and versus electron density derived from
\ion{C}{2} column density using \ion{S}{2} as a proxy (bottom, blue).
The systematic errors due to the conversion from $N$(\ion{S}{2})
(solid red lines) extend beyond the random \ion{S}{2} fitting errors.
Arrows associated with filled symbols indicate errors than extend
beyond the scope of the plot, whereas arrows associated with open
symbols indicate upper limits.\label{fig:cne}}
\end{figure}

\begin{figure}
\epsscale{.8}
\plotone{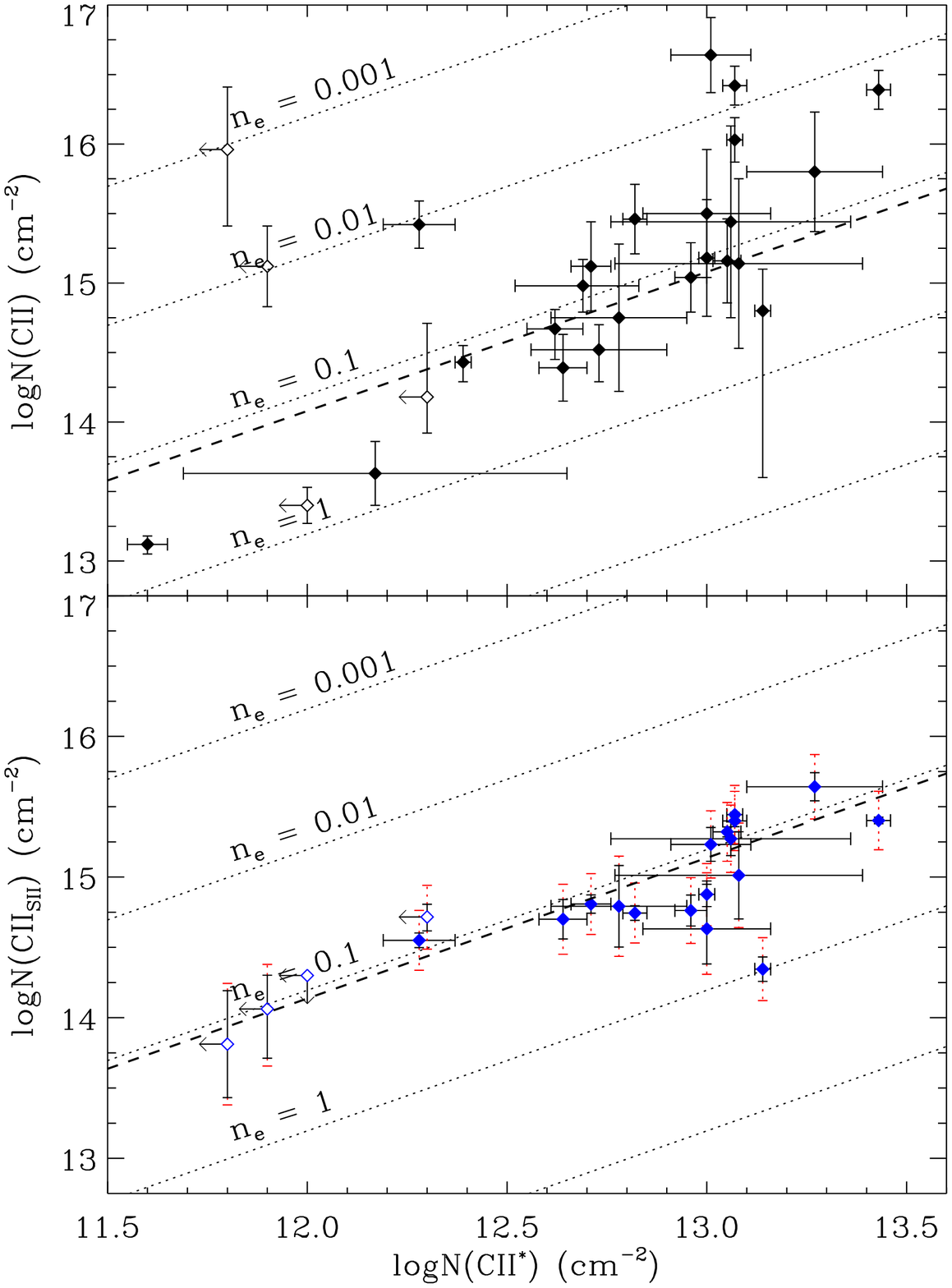}
\caption{Excited line column density versus resonance line column
density (top, black) and versus \ion{C}{2} column density using
\ion{S}{2} as a proxy (bottom, blue).  The systematic errors due to
the conversion from $N$(\ion{S}{2}) (solid red lines) extend beyond
the random \ion{S}{2} fitting errors.  Arrows associated with open
symbols indicate upper limits.  The dotted black lines indicate lines
of equal electron density, assuming the LISM average temperature of
6680~K \citep{redfield04tt}.  The unweighted mean electron densities
are $n_e($\ion{C}{2}$) = 0.13^{+0.15}_{-0.07}$ and
$n_e$(\ion{C}{2}$_{\rm SII}$)$=0.11^{+0.10}_{-0.05}$ (thick dashed
lines), as calculated from the histograms of $n_e$ and
$n_e$(\ion{C}{2}$_{\rm SII}$) in logarithm (See
Figure~\ref{hist}).\label{den2}}
\end{figure}

\begin{figure}
\epsscale{1}
\plotone{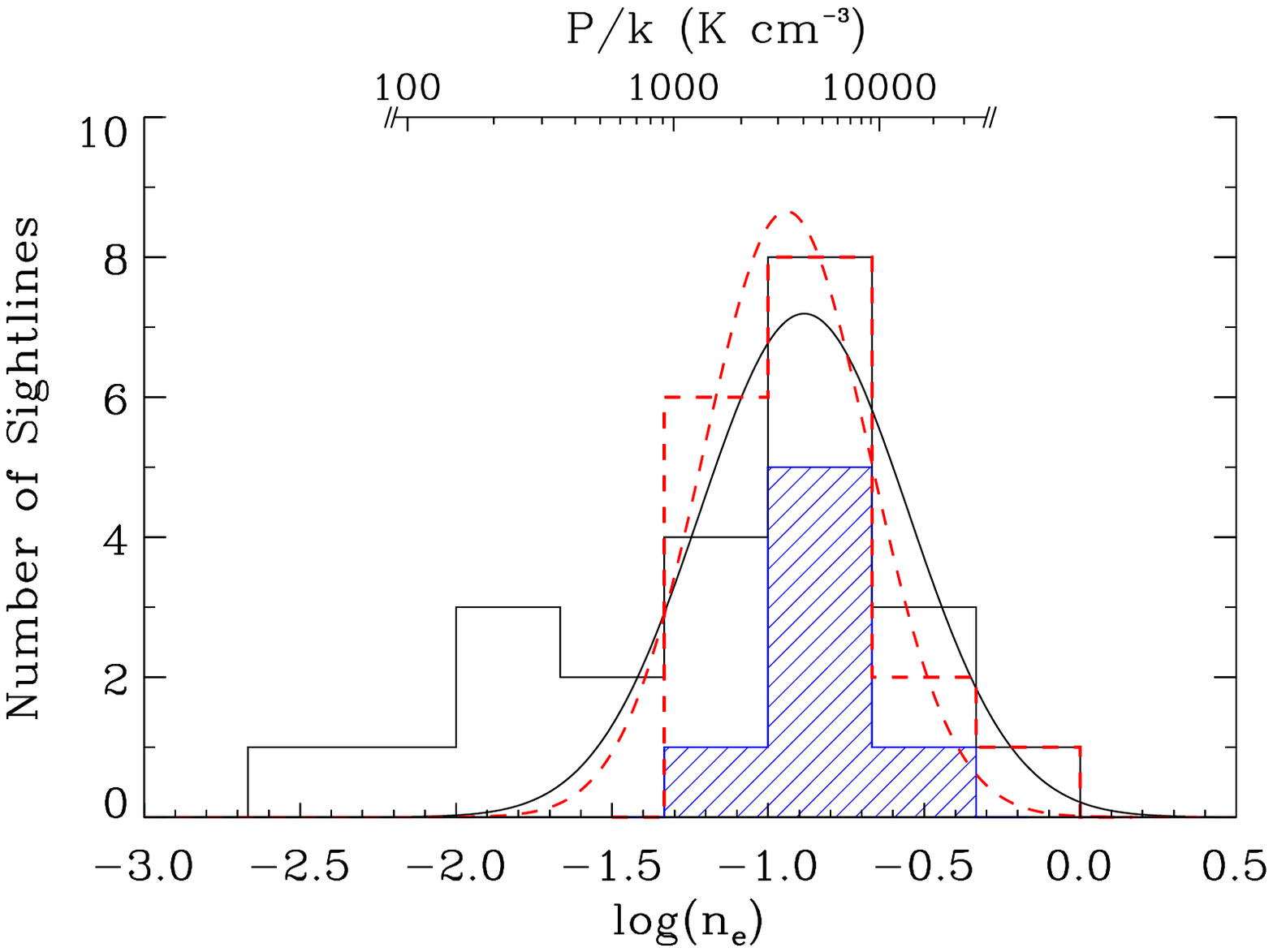}
\caption{Histograms of measured electron densities $n_e$ (solid,
black) and electron densities from \ion{C}{2} column densities based
on \ion{S}{2} as a proxy, $n_e$(\ion{C}{2}$_{\rm SII}$) (dashed, red)
in logarithm with Gaussian fits.  The bin size equals 0.333 dex.  The
unweighted centroid of the solid (black) log-normal distribution is
$-0.88~$log~(cm$^{-3}$) with a dispersion of $0.33~$log~(cm$^{-3}$)
and of the dashed (red) distribution, $-0.94~$log~(cm$^{-3}$) with a
dispersion of $0.26~$log~(cm$^{-3}$).  The shaded (blue) histogram
indicates the electron densities of sight lines that are kinematically
and spatially identified with the LIC.  All LIC sight lines show a
consistent value of $n_e$.  The top axis gives the estimated pressure
$P/k = nT$, assuming the LISM average value of temperature and a
simple photoionization relationship between the electron density
($n_e$) and the neutral hydrogen density ($n_{\rm HI}$).  The axis is
not printed for densities in which we expect these calculations to
fail.  For very low densities, the calculate hydrogen density is lower
than the minimum allowed hydrogen density based on typical observed
hydrogen column densities and distances to the background star.  At
very high densities, given a characteristic size of a LISM cloud, the
resulting high column density ($N_{\rm HI} \geq 10^{19.5}$), will
significantly alter the ionization rate due to shielding and therefore
contradict the assumption of a constant ionization rate.  As shown
here, the distribution in pressure is simply a function of the
distribution of the observed electron density.
\label{hist}}
\end{figure}

\begin{figure}
\epsscale{.7}
\plotone{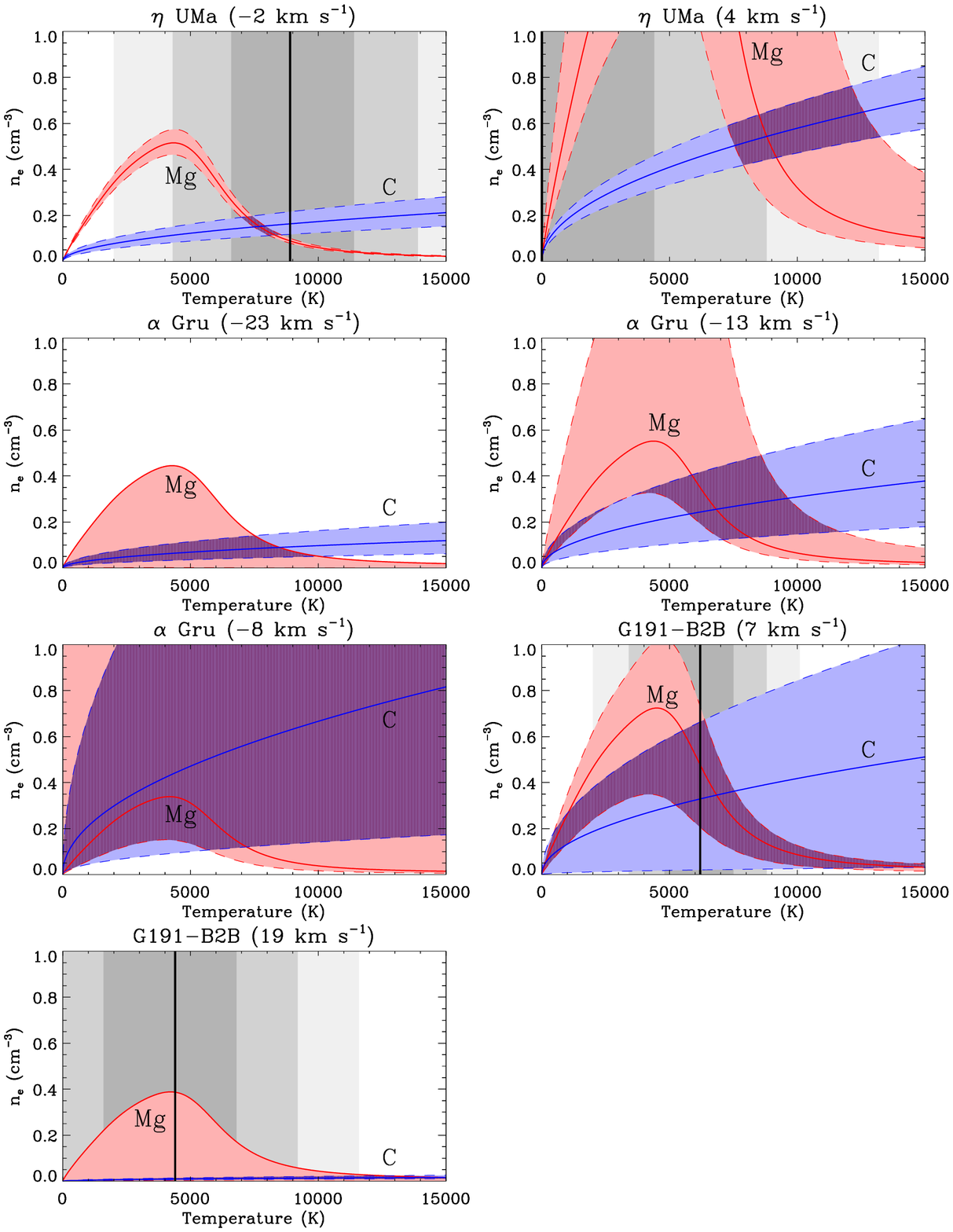}
\caption{Comparison of the temperature dependence in the electron
density calculation using \ion{C}{2}$^{\ast}$ and \ion{C}{2} (blue)
versus \ion{Mg}{1} and \ion{Mg}{2} (red).  This kind of plot was used
by \citet{gry01} to put limits on the temperature and electron density
of clouds along the line of sight toward $\epsilon$ CMa.  However, an
independent measure of the temperature of LISM clouds is available
from comparisons of the line widths of ions of different atomic mass.
\citet{redfield04tt} derive LISM cloud temperatures using this
technique and the temperature for specific components are shown above
by the solid vertical line.  The gray scale above shows the 1$\sigma$,
2$\sigma$, and 3$\sigma$ levels of the temperature.  By combining the
information in line widths and ionization abundance ratios, we can
more tightly constrain both the temperature and electron density of
clouds in the LISM. \label{fig:comp}}
\end{figure}

\begin{figure}
\epsscale{1.15}
\plottwo{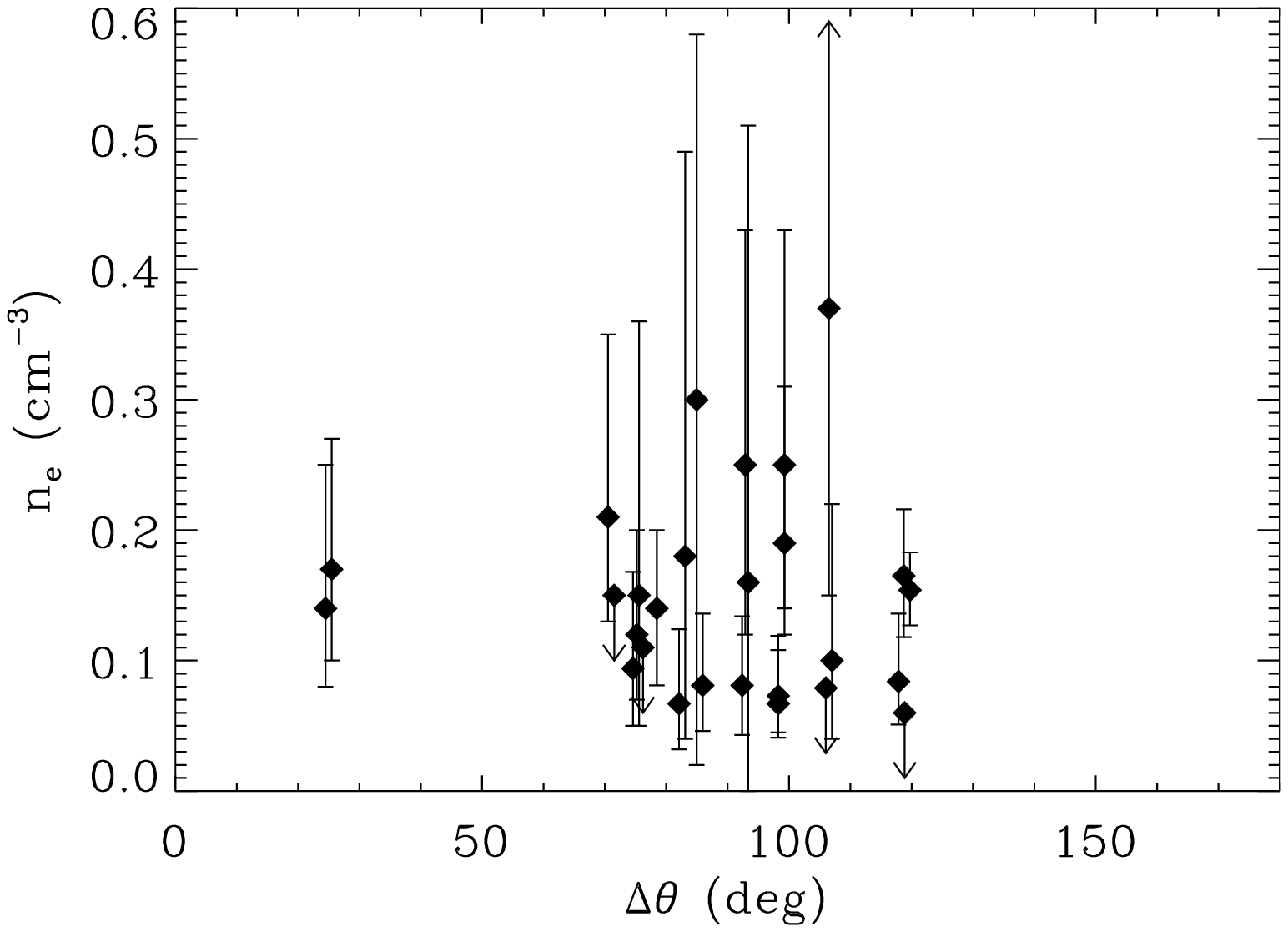}{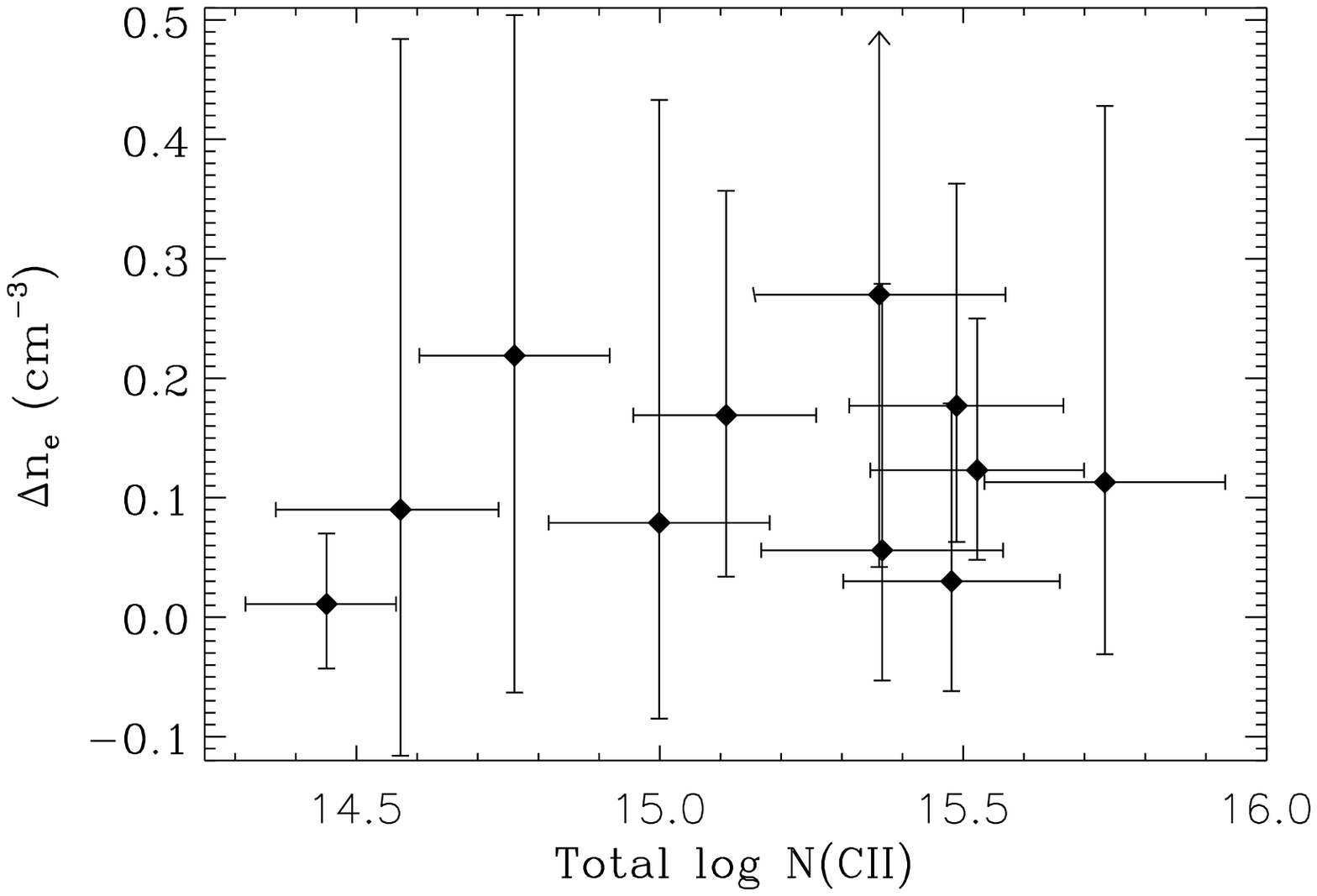}
\caption{{\it left:} Measured electron density as a function of the
angle from the strongest EUV radiation source in the local
environment, $\epsilon$ CMa.  No clear correlation is detected, which
indicates that the ionization structure of the LISM is not solely
determined by this dominant source.  Instead, it is likely that
several of the strong radiation sources in the LISM contribute to the
distribution of electron density measurements.  {\it right:} For sight
lines in which multiple electron densities are measured, we compare
the difference in electron density as a function of total \ion{C}{2}
column density.  No correlation is apparent, which argues that little
shielding of ionizing radiation is occuring among the bulk of these
clouds.  However, there are several examples where we can identify
dynamically the relative distances of multiple clouds along the line
of sight, and the more distant cloud (i.e., the cloud closer to the
ionizing source) has a higher electron density and may shield the
nearer cloud, which subsequently has a lower measured electron
density.
\label{fig:shield}}
\end{figure}

\end{document}